\documentclass[preprint,review,11pt,authoryear]{elsarticle}

\usepackage{mathptmx}
\usepackage{times}
\usepackage{multirow, balance}
\usepackage[vertfit]{breakurl}
\usepackage{graphicx}
\usepackage{caption}
\usepackage{subcaption}
\graphicspath{ {images/} }
\usepackage{booktabs}
\usepackage{multicol}
\usepackage{amsmath}
\usepackage{amssymb}
\usepackage{gensymb}
\usepackage{makecell}
\providecommand{\e}[1]{\ensuremath{\times 10^{#1}}}
\usepackage[font=scriptsize]{caption}
\usepackage{apalike}
\usepackage{pdflscape}
\usepackage{color}

\makeatletter
\long\def\pprintMaketitle{\clearpage
  \iflongmktitle\if@twocolumn\let\columnwidth=\textwidth\fi\fi
  \resetTitleCounters
  \def\baselinestretch{1}%
  \printFirstPageNotes
  \begin{center}%
 \thispagestyle{pprintTitle}%
 \def\baselinestretch{1}%
    \Large\@title\par\vskip46pt
    \normalsize\elsauthors\par\vskip18pt
    \footnotesize\itshape\elsaddress\par\vskip36pt
    \ifvoid\absbox\else\unvbox\absbox\par\vskip10pt\fi
    \ifvoid\keybox\else\unvbox\keybox\par\vskip10pt\fi
    \end{center}%
  \gdef\thefootnote{\arabic{footnote}}%
  }
\makeatother

\journal{Expert Systems with Applications}

\begin{document}

\begin{frontmatter}
\title{\huge Retinal Vessel Segmentation based on Fully Convolutional Neural Networks}

\author[add]{\Large Am\'erico Filipe Moreira Oliveira*}
\author[add]{S\'ergio Rafael Mano Pereira}
\author[add]{Carlos Alberto Batista Silva*}

\address[add]{\large\textnormal{CMEMS-UMinho Research Unit, University of Minho, Campus Azur\'{e}m, Guimar\~{a}es, Portugal.}}
\end{frontmatter}

{\noindent\large Email adresses:

\vspace{0.5cm}

Am\'erico Oliveira: a68396@alunos.uminho.pt

S\'ergio Pereira: id5692@alunos.uminho.pt

Carlos A. Silva: csilva@dei.uminho.pt}

\vspace{1cm}

{\noindent\large *Corresponding authors:

\vspace{0.5cm}

\noindent Am\'erico Filipe Moreira Oliveira;  Carlos Alberto Batista Silva.\newline [Tel.: +351 253510190, Fax: +351 253510189.] \newline
CMEMS-UMinho Research Unit, University of Minho, Dpt. of Industrial Electronics, Campus de Azur\'em, Alameda da Universidade, 4804-533 - Guimar\~aes - Portugal
}

\newpage

\section*{Abstract}

The retinal vascular condition is a reliable biomarker of several ophthalmologic and cardiovascular diseases, so automatic vessel segmentation may be crucial to diagnose and monitor them. In this paper, we propose a novel method that combines the multiscale analysis provided by the Stationary Wavelet Transform with a multiscale Fully Convolutional Neural Network to cope with the varying width and direction of the vessel structure in the retina. Our proposal uses rotation operations as the basis of a joint strategy for both data augmentation and prediction, which allows us to explore the information learned during training to refine the segmentation. The method was evaluated on three publicly available databases, achieving an average accuracy of 0.9576, 0.9694, and 0.9653, and average area under the ROC curve of 0.9821, 0.9905, and 0.9855 on the DRIVE, STARE, and CHASE\textunderscore DB1 databases, respectively. It also appears to be robust to the training set and to the inter-rater variability, which shows its potential for real-world applications.

\vspace{1cm}

\noindent\textbf{Keywords:} Fully Convolutional Neural Network; Stationary Wavelet Transform; Retinal Fundus Image; Vessel Segmentation; Deep Learning.

\newpage

\section{Introduction}

The retinal vascular tree is the only structure in the human circulatory system that can be directly and non-invasively observed \textit{in vivo}, besides being easily photographed \citep{patton}. Apart from other applications, as biometric identification, multimodal image registration, or computer-assisted laser surgery  \citep{fraz_review}, retinal fundus imaging has become essential for diagnosing, treating, and monitoring several disorders,  such as arteriosclerosis, hypertension, age-related macular degeneration, diabetic retinopathy, and glaucoma \citep{abramoff_review, fraz_review}.

The morphological characteristics of the retinal vasculature, such as angles, branching patterns, length, width, and tortuosity, can play a crucial role in assisting cardiologists and ophthalmologists \citep{abramoff_review, fraz_review}. Their quantitative evaluation, however, demands proper segmentation of the vascular tree.

Manually segmenting blood vessels in retinal images is both error-prone and time-consuming, even for experienced physicians. For this reason, automatic and accurate segmentation is vital. This process configures a complex task, not only due to abrupt variations in the attributes (size, shape, intensity levels) and arrangement (branching, crossing) of the vessels but also to the low quality of retinal images \citep{abramoff_review, fraz_review}. If lesions occur, the task becomes even more challenging.

\subsection{Related Work}
\label{sec:RW}

Several works have been proposed for retinal vessel segmentation in recent decades. At large, all of them may be seen as supervised or unsupervised methods.

\newpage

Unsupervised methods make use of prior knowledge on the structure of the vessels and usually rely on rule-based schemes. These algorithms often apply matched filtering, morphological processing, vessel tracking, multiscale, or model-based approaches. Matched filtering techniques, as adopted by  \citet{hoover}, convolve retinal images with a 2D filter to produce a Gaussian intensity profile of blood vessels. \citet{azzopardi} have employed a combination of shifted filter responses (COSFIRE) to detect retinal vessels or any vessel-like pattern. In \citep{zhang}, a 2D image is lifted to a 3D orientation score, and vessels are then enhanced by multiscale derivatives. Present-day approaches seem to reach promising results when dealing with complicated vessel geometries. Still, they may be penalized by the unbalanced filter response to pathological vessels. Strategies based on mathematical morphology combined with matched filtering for centerline detection, as in \citep{mendonça}, or adaptive thresholding, as in  \citep{roychowdhury2015, neto2017}, can also be found. Vessel tracking techniques, such as those of \citet{liu} and \citet {yintracking}, establish an initial set of points (selected either manually or automatically) and iteratively follow the vessel centerlines to extract the vascular tree. However, besides being poor at detecting not seeded vessel segments, they can miss initially flagged ones too. Multiscale techniques, like those of  \citet{trucco} and  \citet{nguyen}, try to improve the analysis of blood vessels of varying width by adjusting the scale to the diameter of the vessel, but selecting the parameters of a multiscale filter is not a trivial task. Finally, model-based techniques either apply explicit vessel profile or deformable models to extract the vasculature.  \citet{lam} came up with a multi-concavity method to treat, simultaneously, both healthy and pathological images. \citet{zhao} showed an infinite active contour model that combines intensity and local phase information. In general, unsupervised methods may generalize well across images, since they are not learned from a sample of the population as the supervised ones. Still, it may be difficult to encompass both normal and abnormal vasculatures due to their diverse appearing.

Supervised methods use ground truth data to train a classifier to label each pixel as either vessel or background. During the training stage, an optimization algorithm analyses the desired output for each training example and infers a function, which the classifier uses for dealing with unseen examples, in the testing stage.  \citet{niemeijer} started by introducing a vector of features, based on Gaussian functions and their derivatives, for each pixel, and then applied a k-nearest neighbor classifier. Later, \citet{staal} extended this method by introducing ridge profiles in vessel detection. \citet{soares} used a gaussian mixture model classifier and computed a 6-feature set extracted by a Gabor wavelet transform. \citet{marin} applied neural networks to classify a 7-feature set composed of gray-level and moment invariant-based features. \citet{fraz} combined the analysis of the gradient vector, line strength measures, and Gabor filter responses to compute a 9-feature set, which was passed to an AdaBoost classifier. \citet{roychowdhury2014}, in turn, used a gaussian mixture model to classify a 8-feature set based on each pixel's neighborhood and first and second-order gradient images. \citet{strisciuglio2016} came up with a bank of COSFIRE filters and trained a support vector machine classifier to determine the most effective subset of filters. \citet{orlando} presented a fully connected conditional random field model, whose parameters were learned with a structured output support vector machine. Recently,  \citet{zhang2017} employed a random forest classifier to deal with a 29-feature set built by combining vessel filtering and wavelet transform features. Contrasting with the previous unsupervised approaches, supervised methods can bypass the need for designing different models whenever new segmentation problems arise. Still, results are often strongly affected by the feature engineering stage.

Deep learning-based methods have recently drawn attention since they can automatically learn an increasingly complex hierarchy of features directly from the input data \citep{bengio}. This hints a paradigm shift according to which people are now optimizing architectures, instead of designing hand-crafted features that may be problem dependent and require expert knowledge. Even if we can trace deep neural networks back to the last century \citep{lecun1990}, they have recently been applied to several areas, including retinal vessel segmentation. \citet{feng}, for example, turned the segmentation task into a vessel mapping  problem, where the mapping function was learned by a 5-layer deep neural network. In general, this recent boost is due to the fortunate combination of two factors: (1) the amount of data available for training is now massively larger than it was decades ago; and (2) it coincided with the development of more powerful graphical processing units \citep{lecun}. 

Convolutional neural networks (CNNs), in particular, have already been used to win quite a few object recognition \citep{krizhevsky, dieleman} and biological image segmentation \citep{ronneberger} challenges. In retinal image analysis, recent proposals also employ them. \citet{melinscak} addressed vessel segmentation using a 10-layer CNN. In \citep{liskowski}, a structured prediction scheme was used to highlight context information, while testing a comprehensive set of architectures among which a 7-layer no-pooling CNN was the most successful. Recently, \citet{fu} combined a typical 7-layer CNN with a conditional random field, reformulated as a recurrent neural network, to model long-range pixel interactions.

While previous works on deep learning used only raw data, which was processed by cascaded convolutional layers to learn features, it has been recently suggested that deep neural networks can generate more useful features when provided with extra information. This has been, for instance, explored using wavelets for SAR image segmentation \citep{duansar} or in super-resolution \citep{guo}. We further consider the use of wavelets,  combined with a multiscale CNN, for segmenting retinal vessels.

\subsection{Motivation and Contributions}

In a preliminary version of this work \citep{oliveira}, we presented a CNN that achieved promising results in retinal vessel segmentation. This paper extends our earlier work. The Stationary Wavelet Transform (SWT) was incorporated to exploit the multiscale nature of the vascular system. An alternative data augmentation strategy was used, and a new multiple prediction scheme, motivated by it, was introduced. Finally, additional ablation studies and implementation details better cover the underlying principles of our method. 

The main idea of this approach is to combine information with different levels of abstraction. In a typical CNN, several pooling operators successively reduce the dimensions of the feature maps that are being produced by each convolutional layer. This builds high-level information maps, with low resolution. Although these maps contain more complex and compact features, some details may be lost. Thus, the low-level information available in the shallow layers may help. In other words, we provide detail information to the network, which must learn how to use it, to mitigate the losses of information that may occur during the traditional path.

The main contributions of this paper to the retinal vessel segmentation problem are as follows.  We propose rotation operations as the basis of a joint strategy for data augmentation and prediction. Although data augmentation is a well known technique, here we explore the information learned during training, about the arrangement and orientation of the vessels, to refine the segmentation. We also investigate the decomposition of the image through SWT as a way to add new input channels into a Fully Convolutional Neural Network (FCN); this contribution is independent of the architecture, but here we show how to combine a multiscale architecture with the SWT for better handling retinal vessels at different scales.

Our method was evaluated on three publicly available databases: DRIVE \citep{staal}, STARE \citep{hoover}, and CHASE\textunderscore DB1 \citep{owen}. A worth mentioning methodological contribution is the comparison between the annotations provided by different physicians and the study of the effects that this inter-rater variability can have on the behavior of the network.

The remaining sections are organized as follows. The proposed method is described in section \ref{sec:method}. The experimental setup is presented in section \ref{sec:experimental}. Results and discussion can be found in section \ref{sec:results}.  Before closing, we come up with the main conclusions in section \ref{sec:conclusions}.

\begin{landscape}
\centering
\begin{figure*}[t!]
\centering
    
\makebox[\columnwidth]{
\begin{subfigure}[h]{1.6\textwidth}
   \includegraphics[width=\textwidth]{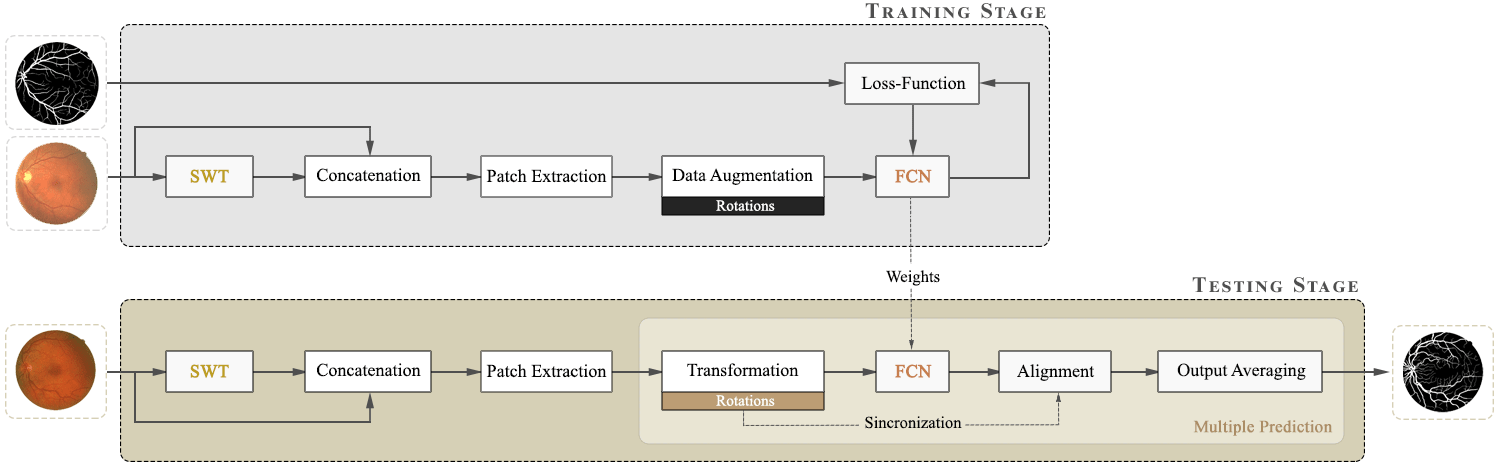}
   \subcaption{}
   \label{fig:overa} 
   \vspace{0.1cm}
\end{subfigure}}

\makebox[\columnwidth]{
\begin{subfigure}[h]{1.6\textwidth}
   \includegraphics[width=1\textwidth]{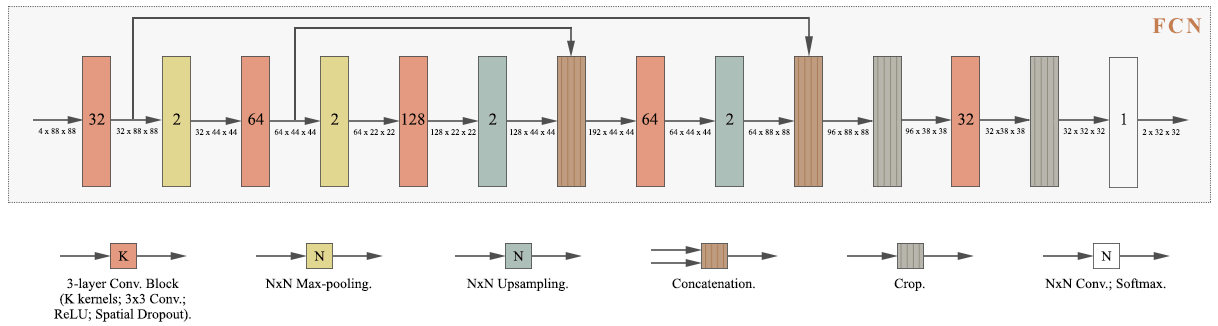}
   \subcaption{}
   \label{fig:overb}
\end{subfigure}}

\caption{Overview of the proposed method: (a) Block diagram; (b) FCN architecture.}
\end{figure*}
\end{landscape}

\section{Method}
\label{sec:method}

Fig. \ref{fig:overa} presents an overview of the proposed approach. There are four main stages: input building through the SWT, patch extraction, classification via FCN, and multiple prediction.

\subsection{Stationary Wavelet Transform}
\label{sec:swt}

The SWT \citep{holschneider} was initially designed to overcome two drawbacks of the discrete wavelet transform (DWT): (1) DWT is not translation-invariant; and (2) it can only be used on images of dyadic size \citep{holschneider}. Here we propose the SWT as a method to enrich the input of the FCN. We employed the SWT over the DWT since it does not downsample the coefficients, preserving the initial number of pixels. Thus, it allows us to add new extra channels to the input. 

\begin{figure}[t!]
\centering
   \begin{subfigure}[h]{1\textwidth}
   \includegraphics[width=1\textwidth]{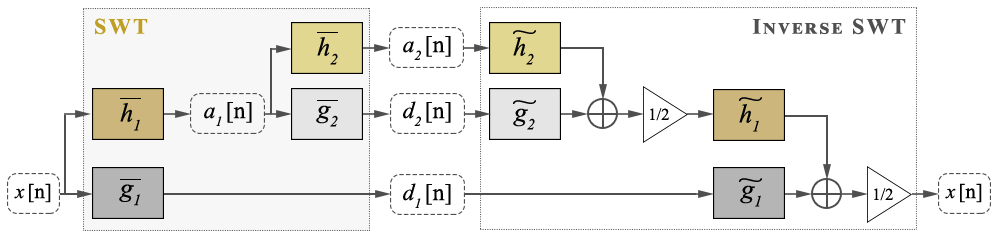}
   \subcaption{}
   \label{fig:swta} 
\end{subfigure}
\begin{subfigure}[h]{1\textwidth}
   \includegraphics[width=1\textwidth]{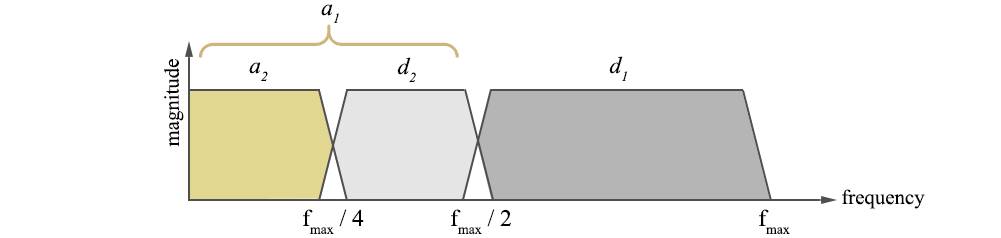}
   \subcaption{}
   \label{fig:swtb}
\end{subfigure}
\caption{SWT operation: (a) Filter bank; (b) Spectral analysis.}
\label{fig:swt}
\end{figure}


Fig. \ref{fig:swt} gives some insights on how SWT operates. For simplicity's sake, we show a 2-level decomposition with the 1D transform. $\overline{g_j}$ and $\overline{h_j}$ are, respectively, the high-pass and low-pass filters of each level $j$. The former returns the detail coefficient ${d_j}$; the latter transfers the approximation coefficient ${a_j}$ to the level $j+1$ for further decomposition (Fig. \ref{fig:swta}). This filter pair splits the input signal into two spectral bands (Fig. \ref{fig:swtb}). At each level $j+1$, the impulse response of each low-pass filter is obtained by upsampling the corresponding response of the previous level $j$ \citep{holschneider}:

\begin{equation}
\overline{h_{j+1}} [k] = \overline{h_j} [k] \uparrow = \begin{cases} \overline{h_j} [\frac{k}{2}] & k\mbox{ even}, \\ 0 & k\mbox{ odd}, \end{cases}
\end{equation}

\begin{figure}[!t]
		\centering
		\begin{multicols}{4}
		\includegraphics[width=.26\textwidth]{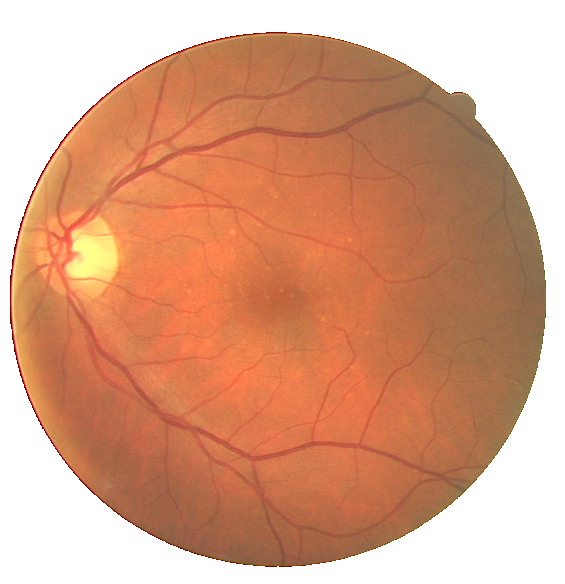}   \par\subcaption{}\label{fig:image}
		\includegraphics[width=.27\textwidth]{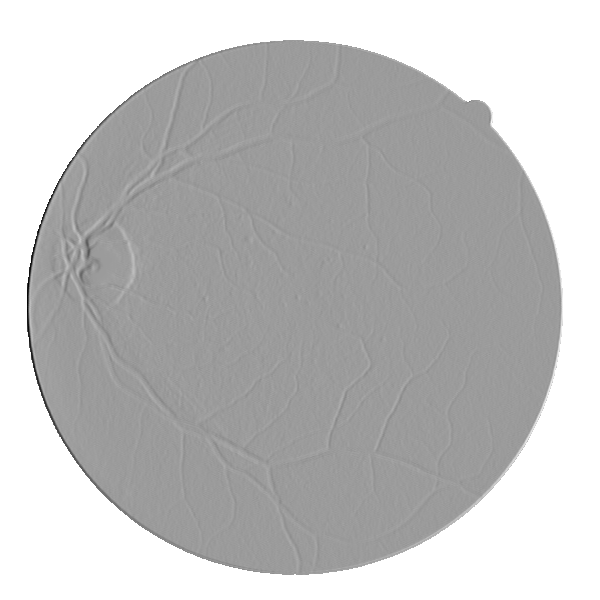}   \par\subcaption{}\label{fig:dV1}
		\includegraphics[width=.27\textwidth]{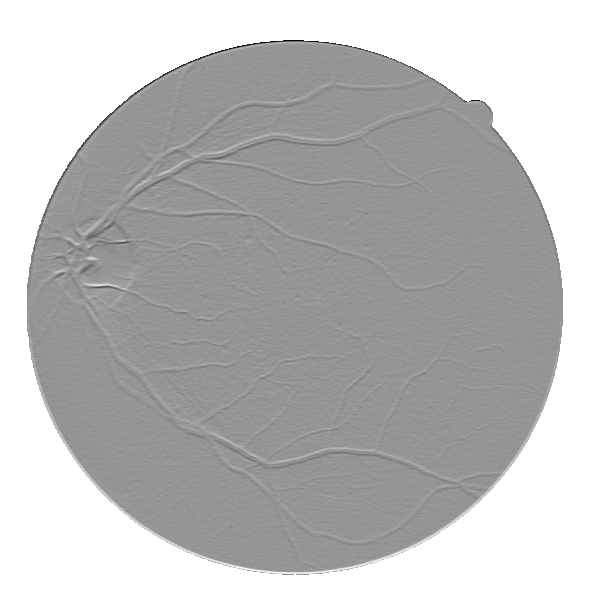}  \par\subcaption{}\label{fig:dH1}
		\includegraphics[width=.27\textwidth]{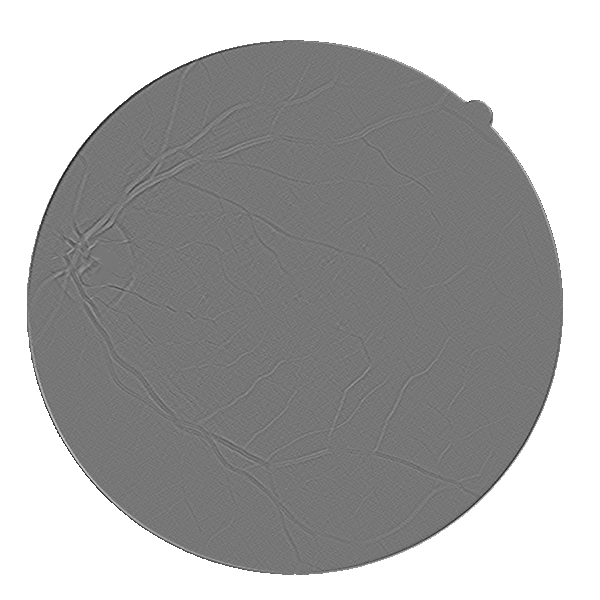}  \par\subcaption{}\label{fig:dD1}
		\end{multicols}
		\vspace{-1cm}
		\begin{multicols}{4}
		\includegraphics[width=.26\textwidth]{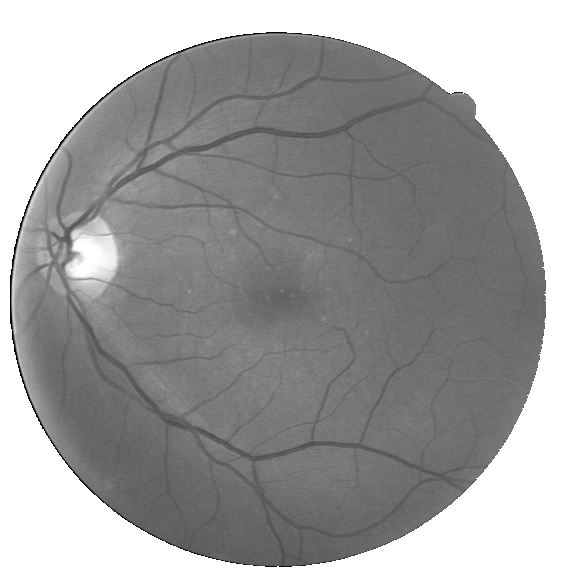}   \par\subcaption{}\label{fig:green}
		\includegraphics[width=.27\textwidth]{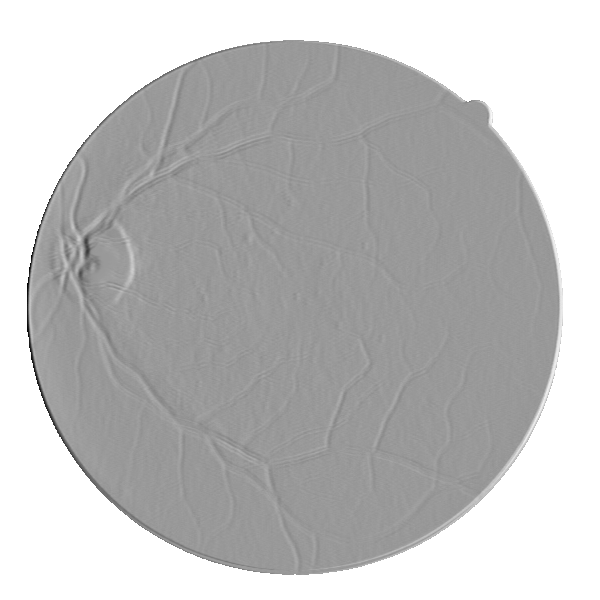}   \par\subcaption{}\label{fig:cV2}
		\includegraphics[width=.27\textwidth]{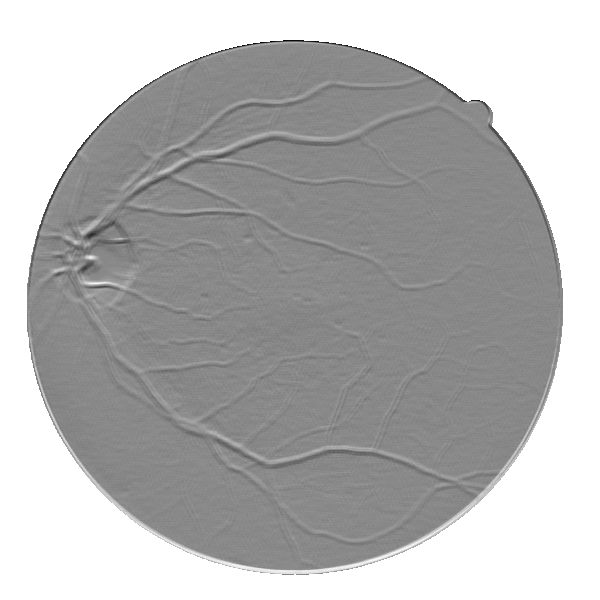}  \par\subcaption{}\label{fig:cH2}
		\includegraphics[width=.27\textwidth]{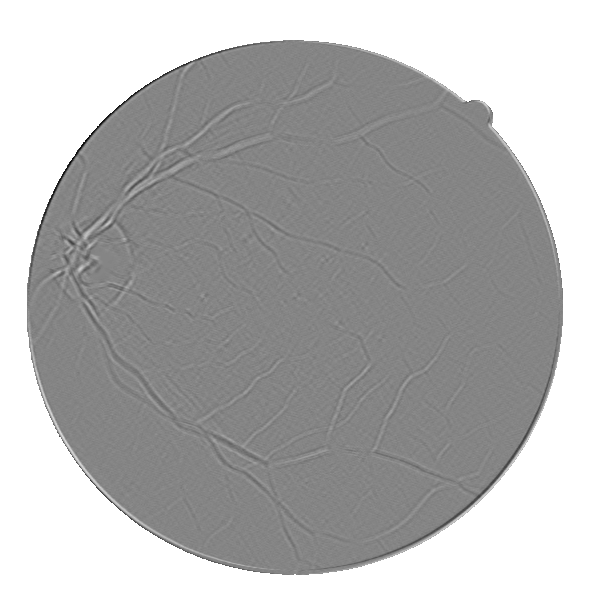}  \par\subcaption{}\label{fig:cD2}
		\end{multicols}
		\vspace{-0.2cm}
		\caption {Application of the SWT to a retinal image: (a) Color image; (b) $dV_\textrm{1}$; (c) $dH_\textrm{1}$; (d) $dD_\textrm{1}$; (e) Green channel; (f) $dV_\textrm{2}$; (g) $dH_\textrm{2}$; (h) $dD_\textrm{2}$.}
		\label{fig:swtchannels}
	\end{figure}

\vspace{0.2cm}

\noindent with the equivalent equation being applied to the high-pass case. Each signal of level $j+1$ is obtained by convolving the corresponding signal of the previous level $j$ with $\overline{g_j}$ or $\overline{h_j}$. These signals keep the original dimensions of the input and are translation invariant. Moreover, they contain multi-resolution information, which is very useful for segmenting images. The reconstruction filters $\widetilde{g_j}$ and $\widetilde{h_j}$ are time-reversed versions of their homologous decomposition filters. When dealing with images, the 2D transform is required. This implies vertical, horizontal, and diagonal versions of the process, resulting in three detail images (${dV_j}$, ${dH_j}$, and ${dD_j}$) at each level $j$  (Fig. \ref{fig:swtchannels}).

 Throughout this work, we used the Haar wavelet \citep{haar1910}, which can be described as: 

\begin{equation}
\psi(t) = 
\begin{cases} 
1 & 0 \leq t < \frac{1}{2}, \\ 
-1 & \frac{1}{2} \leq t < 1, \\
0 & otherwise.
\end{cases}
\end{equation}

\vspace{-0.2cm}

\subsection{Patch Extraction}
\label{sec:patchext}

In our approach, the patches fed to the network were obtained differently according to the stage of the algorithm. During training, we extracted 2750, 3250, and 3750 patches from each image of the DRIVE, STARE, and CHASE\textunderscore DB1 databases, respectively. We noticed that the network benefited from using more patches in larger images, and these values were experimentally found. In addition, we did not apply any restrictions at this stage, so overlapping patches were allowed. During testing, however, we ensured there was no overlap between the output patches, i.e., each pixel was segmented only once. This was done by zero padding the original image in such a way that its dimension becomes an integer multiple of the output patch size. Furthermore, we applied an overlap-tile strategy \citep{ronneberger}, where each output patch only contains the pixels for which the full context is available in the input image.  This was the main reason for the cropping operations, which led to the mismatch of dimensionality between the $88\times88$ input patch and the $32\times32$ output patch (Fig. \ref{fig:overb}).

\subsection{Fully Convolutional Neural Network}
\label{sec:network}

CNNs are based on convolutional layers, which receive a stack of input planes and return a group of feature maps. Each feature map results from convolving a single kernel over the inputs, with each layer having a variable number of kernels. 

If we denote the $k$th kernel by $W_k$, the computation of the associated feature map $F_k$ can be formally described as:

\begin{equation}
F_k = b_k + \sum_{c} (W_k){_c} * X_c\,,
\end{equation}

\noindent where $(W_k){_c}$ and $X_c$ are, respectively, the $c$th channel of both the kernel and the input, \text{*} is the convolution operation, and $b_k$ represents the bias term.

From the neural network perspective, each unit of a feature map is the output of a neuron that captures particular features from a restricted region of the previous layer. That region -- called the \textit{receptive field} of the neuron -- is defined by the kernel size. If a unit has a high value, then the kernel has found a match and the feature encoded by it (like an edge or a curve) is, at least partially, present in the input. Otherwise, there is a less relevant correspondence. As new layers are stacked, and deeper nets appear, more complex features (as motifs, parts, or objects) are recognized.

Among the reasons that make CNNs well-suited to processing visual information, there are two key ideas: local connections and shared weights. Local connectivity means that each hidden unit only looks for its own receptive field, which significantly reduces the number of weights. Weight sharing happens since the same set of weights is convolved over the whole image, which also increases computational efficiency and provides CNNs with translation invariance \citep{lecun}.

In the following lines, we address our decisions regarding some fundamental aspects when dealing with CNNs.

\subsubsection{Initialization} We adopted the Xavier initialization \citep{xavier}, which allowed us to maintain the gradients in controlled levels and thus prevent gradient vanishing during back-propagation.

\subsubsection{Activation Function} Herein, our choice fell on the rectified linear units (ReLU) \citep{nair}, $f(x) = \max(0, x)$, which were found to speed up training in comparison with other  nonlinearities, such as the sigmoid or the hyperbolic tangent functions \citep{krizhevsky}.

\subsubsection{Pooling}
We employed max-pooling, which discards possibly redundant features, making the representation invariant to small details \citep{lecun}.

\subsubsection{Upsampling}
\label{sec:upsampling}
To return the feature maps to their initial dimensions, we employed nearest neighbor interpolation.

\subsubsection{Regularization}
\label{sec:reg}

\citet{tompson} suggested that when facing natural images with high spatial correlation, standard dropout \citep{srivastava} may be less efficient since neighboring units in the same feature map also become heavily correlated. Thus, we adopted spatial dropout \citep{tompson}, which removes entire feature maps instead of just some nodes. In this way, adjacent units in each feature map are either all neutralized or all active. The same probability $p$ was used in all convolutional blocks, except in the last one where we applied $p_{last}$.

\subsubsection{Architecture}
\label{sec:arch}

In classic image recognition CNNs \citep{krizhevsky, simonyan}, it was common to identify a feature extraction stage based on the convolutional and pooling layers, and a classification stage carried out by a variable number of fully connected (FC) layers. The problem with this type of networks, which we will refer to as FC-CNNs, was that they were forced to take fixed-sized inputs since they had at least one FC layer. In FCNs \citep{long}, all the FC layers are replaced by  convolutions (often with $1\times 1$ kernels). In this way, a FCN can take an image, of any size, as input, and directly output a set of probability maps with the same dimensions. This, besides bringing advantages regarding versatility, makes FCNs more  efficient than FC-CNNs, since it is possible to segment a set of pixels at once, instead of treating each pixel singly.

With the appearance of FCNs, multi-scale architectures, also known as encoder-decoder networks, became more popular. The work by Long et al. \citep{long} was perhaps the first to investigate the idea of merging feature maps with different levels of abstraction, but several other multi-scale methods for semantic segmentation followed in a short time \citep{segnet,boxsup,predicting,hypercolumns,learning}. In medical image segmentation, this trend led to the U-net \citep{ronneberger}.

Our architecture, which is shown in Fig. \ref{fig:overb}, is inspired by these previous works. The network is composed of both an encoder and a decoder. In the former, pooling is applied to summarize neighboring features and create high-level representations. In the latter, feature maps are reconstructed, combining those coming from the encoder, through skip connections, with those belonging to low-level scales. While pooling operations are highly suitable for recognizing objects in images and crucial for increasing the receptive field of deeper units, they can significantly harm localization performance. In fact, we seem to face an intriguing problem: high-level information reveals \textit{what}, but low-level information reveals \textit{where} \citep{long}. In other words, building a one-way path capable of achieving satisfactory results may not be feasible. This is the main motivation for the decoder structure since skip connections allow information to propagate directly from shallow, high-resolution, low-level information layers to deep, low-resolution, high-level information ones. Another important aspect is that after each max-pooling in the encoder path we double the number of feature maps, whereas in the decoder path we halve it after each upsampling. This is mainly due to computational reasons, since larger feature maps lead to higher computational load.

In our network, the first channel of the 4-channel input patch resulted from taking the green channel of the retinal image and normalizing it to have zero mean and unit variance. The remaining channels were obtained by applying the SWT over the first and then subjecting them to the same normalization procedure. In section \ref{sec:base}, we evaluate our proposal using just the green channel. In section \ref{sec:rswt}, we show the benefits of including the SWT channels. We used small $3 \times 3$ kernels in all convolutional layers (except in the  $1 \times 1$ linear convolution). Stacking smaller kernels can ensure the same effective receptive field of bigger ones while reducing the number of weights \citep{simonyan}. These and other hyper-parameters are summarized in Table \ref{tab:param}. Therein, a mini-batch size of 4 means that four input patches were propagated through the network in each iteration.

\begin{table}[h]
\caption{Hyperparameters of the proposed FCN}
\begin{center}
\renewcommand{\arraystretch}{1}
\scalebox{0.9}{
\begin{tabular}{ccc}
\specialrule{.2em}{.1em}{.1em}
\multicolumn{1}{c}{\textbf{Stage}} & \multicolumn{1}{l}{\textbf{Hyperparameter}} & \multicolumn{1}{r}{\textbf{Value}} \\ 
\specialrule{.12em}{.1em}{.1em}
\multirow{2}{*}{Initialization} 
               & \multicolumn{1}{l}{Bias}             &\multicolumn{1}{r}{0.1}     \\ 
               & \multicolumn{1}{l}{Weights}          & \multicolumn{1}{r}{Xavier} \\
\midrule
\multirow{5}{*}{Training} 
               & \multicolumn{1}{l}{Epochs}       & \multicolumn{1}{r}{20}      \\ 
               & \multicolumn{1}{l}{Mini-batch size}           & \multicolumn{1}{r}{4}     \\
               & \multicolumn{1}{l}{Spatial dropout $p$}       & \multicolumn{1}{r}{0.2}      \\ 
               & \multicolumn{1}{l}{Spatial dropout $p_{last}$}       & \multicolumn{1}{r}{0.15} \\ 
               & \multicolumn{1}{l}{Learning rate decay $\lambda$}    & \multicolumn{1}{r}{$1\e{-6}$}    \\ 
\specialrule{.01em}{.1em}{.1em}
\end{tabular}}
\end{center}
\label{tab:param}
\end{table}

\subsubsection{Training}
\label{sec:training}

During the optimization process, we applied stochastic gradient descent with Nesterov momentum \citep{nesterov} to minimize a categorical cross-entropy loss-function:

\begin{equation}
J(w) = - \frac{1}{M} \sum_{j=1}^M \sum_{k=1}^K y_{j,k} \ ln(\hat{y}_{j,k}),
\end{equation}

\noindent where $M$ and $K$ denote, respectively, the number of pixels and classes, $\hat{y}$ refers to the probabilistic predictions (after the softmax), and $y$ is the target.

Furthermore, both the momentum $\nu$ and the learning rate $\eta$ were changed in specific epochs, according to Table \ref{tab:optparam}. Besides that, we still further decayed the latter one in-between these changes, according to:

\begin{equation}
\eta_n = \frac{\eta_{n-1}}{1 + \lambda \times n}\,,
\end{equation}

\noindent where $\eta_n$ and $\eta_{n-1}$ are, respectively, the learning rate in the present and previous updates, and $\lambda$ is the learning rate decay. Herein, an epoch means a complete pass over all the training samples, while an update refers to the update of the weights after each iteration.

\begin{table}[t]
\caption{Schedule of the optimization parameters}
\begin{center}
\renewcommand{\arraystretch}{1}
\scalebox{0.9}{
\begin{tabular}{ccc}
\specialrule{.2em}{.1em}{.1em}
\multicolumn{1}{c}{\textbf{Parameter}} & \multicolumn{1}{c}{\textbf{Epoch}} & \textbf{Value} \\  
\specialrule{.12em}{.1em}{.1em}
\multirow{4}{*}{Learning rate $\eta$}  
              & \multicolumn{1}{r}{1}             &\multicolumn{1}{r}{0.05}     \\ 
              & \multicolumn{1}{r}{10}             &\multicolumn{1}{r}{0.02}     \\ 
              & \multicolumn{1}{r}{14}             &\multicolumn{1}{r}{0.002}     \\ 
              & \multicolumn{1}{r}{18}             &\multicolumn{1}{r}{0.0002}     \\
\midrule
\multirow{3}{*}{Momentum $\nu$} 
              & \multicolumn{1}{r}{1}             &\multicolumn{1}{r}{0.2}     \\ 
              & \multicolumn{1}{r}{10}             &\multicolumn{1}{r}{0.9}     \\ 
              & \multicolumn{1}{r}{14}             &\multicolumn{1}{r}{0.99}     \\ 
\specialrule{.01em}{.1em}{.1em}
\vspace{-0.1cm}
\end{tabular}}
\end{center}
\label{tab:optparam}
\end{table}

\subsection{Data Augmentation and Multiple Prediction}
\label{sec:PW}

In our data augmentation procedure, each original patch was rotated by 90$\degree$, 180$\degree$, and 270$\degree$. Then, these patches were randomly placed in the training vector. The number of patches was increased by four (since the original one was included too). To avoid interpolations, only multiples of 90$\degree$ were used. Also, each operation was applied to both the original patch and the respective annotated one to keep the correspondence.

We restricted augmentation to rotations, as they proved to be particularly useful. Even though rotations are common in data augmentation, here we intend to clarify how to use them in the specific context of FCNs. In fact, many valuable proposals that do data augmentation with rotations, in FC-CNNs, can be found in the literature. A good example is to embed the rotational augmentation in the network itself, as in Dieleman et al. \citep{dielemancyclic} and Worrall et al. \citep{worrall2016harmonic}. However, these strategies were developed for FC-CNNs and cannot be applied in FCNs for two main reasons. First, when dealing with FCNs, there is a correspondence between each pixel in the input patch and in the subsequent feature maps; so, if we rotate the feature maps inside the network, as embedded layers, and further add them, as these works suggest for FC-CNNs, we would lose such correspondence. Second, these strategies require the rotated versions to be arranged consecutively, which we show in section \ref{sec:datacomponents} to be detrimental to the performance.

While for both FC-CNNs and FCNs, a mini-batch contains patches that group together neighbor pixels, a particularity in the training of FCNs is that the computation of the loss takes into account the error of each neighbor pixel in the patch. This seems to indicate that the network will encode context information provided by the combination of the pixels. Considering the positive effects of the rotational augmentation in the generalization capacity of the network, we may conclude that the context provided by the rotations brought new information that was encoded in the network. Although this encoded information was used efficiently in FC-CNNs by works as \citep{dielemancyclic, worrall2016harmonic}, the same was not achieved in FCNs, and we believe this is due to the internal alignment required for a proper use of FCNs.  However, this limitation can be surpassed if we apply the rotations differently during training and prediction. In training, we must present the rotated patches randomly in order to effectively train the network; however, in prediction, they should be presented consecutively to ensure that we are able to fuse the information of all possible contexts. With this in mind, we employed the following procedure (recap Fig. \ref{fig:overa}). We started by generating three artificial rotated copies of each patch to be segmented, as described above, making up a set of four examples -- Transformation. Each example was then segmented and we obtained the respective probability map of each class. Finally, the probability maps of the rotated patches underwent a rotation by the same initial angle, but in the opposite direction -- Alignment. This allowed the four maps of each class to be pixel-wise merged by averaging, building the final unique segmentation -- Output Averaging.

\section{Experimental Setup}
\label{sec:experimental}

\subsection{Materials}

The proposed method was evaluated using three publicly available databases.

The DRIVE database \citep{staal} consists of 40 images, 7 of which showing pathological signals. Each image has a resolution of $565 \times 584$ and 8 bits per channel. The STARE database \citep {hoover} contains 20 images, 10 of which belonging to sick individuals. These images have a resolution of $700 \times 605$ and 24 bits per channel. The CHASE\textunderscore DB1 database \citep{owen} has 28 images, collected from both the eyes of 14 children. Each image has a resolution of $999 \times 960$. 

In the DRIVE case, the global set is divided into the training and testing sets, each of them with 20 images. Thus, the model was evaluated on the testing set. In the remaining two cases, however, there is no clear division. Given this,  the model was trained using a stratified $k$-fold cross-validation, where the original set was partitioned into $k$ equal-sized folds. The validation process was repeated $k$ times, with each fold being retained as testing set, and the remaining $k-1$ folds being used for training. The results were then averaged to produce a single estimation. In the STARE case, we settled $k = 5$, having 5 folds of 4 images. Besides, we stratified the folds by ensuring that half of the images in each of them belong to pathological individuals. In the CHASE\textunderscore DB1 case, we used $k = 4$, obtaining 4 folds of 7 images. This time, each fold included 3 images of one eye and 4 images of the other.

When using the $k$-fold cross-validation, the same architecture was used in all folds. The network was trained from scratch in each fold.

For each database, there are two manual segmentations available made by two independent human observers. For the DRIVE and CHASE\textunderscore DB1 databases, we used the annotations of the first human observer as ground truth. For the STARE database, the gold standard were the annotations from Hoover. The human observer performance was measured using the manual segmentations of the second human observer. The binary masks for DRIVE images are publicly available. For the remaining databases, we have manually created them.

\subsection{Evaluation Metrics}

To enable comparison with other state-of-the-art works, we used four metrics commonly found in the literature: sensitivity (\textit{Sn}), specificity (\textit{Sp}), accuracy (\textit{Acc}), and area under the ROC curve (AUC). The perfect classifier would hit 1 in all of them.

\subsection{Implementation Details}

The proposed method was implemented using Keras\footnote{https://github.com/fchollet/keras} with TensorFlow\footnote{https://github.com/tensorflow/tensorflow} backend and cuDNN 5.1. All tests were conducted on a desktop equipped with a NVIDIA GeForce GTX 1070 GPU, an Intel Core i7-6850K CPU @ 3.60GHz processor, 128 GB of RAM, and running Linux Mint 18 OS.

\section{Results and Discussion}
\label{sec:results}

We begin this section by validating the key components of the \textit{Base System}. Then, we obtain the best model by adding the channels obtained via SWT to it. Having finished these steps, we compare our best model with other state-of-the-art works. The clinical applicability is assessed using a cross-training strategy. Finally, we analyze the behavior of the model when facing the inter-rater variability.

Due to the size of the population and the uncertainty regarding the normality of the data, all tests of statistical significance were computed using the two-sided paired Wilcoxon signed-rank test \citep{wilcoxon} (significance level: 0.05). 

In sections \ref{sec:base} and \ref{sec:rswt}, we performed ablation studies where each component under study was removed or replaced, and the results were recomputed. For this purpose, we used the DRIVE database and focused on \textit{Acc} and AUC. Regarding the tests of statistical significance, we evaluated the following null hypothesis: the results obtained with each variant are not statistically different from those of the \textit{Base System}.

\subsection{Validation of the Base System}
\label{sec:base}

\begin{figure}[!t]
		\centering
		\begin{multicols}{4}
		\includegraphics[width=.25\textwidth]{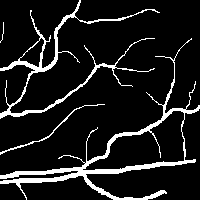}   \par\subcaption{}\label{fig:gt}
		\includegraphics[width=.25\textwidth]{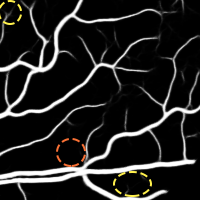}  \par\subcaption{}\label{fig:2750}
		\includegraphics[width=.25\textwidth]{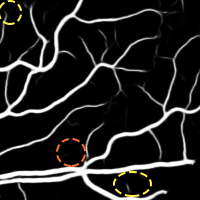}  \par\subcaption{}\label{fig:12000}
		\includegraphics[width=.25\textwidth]{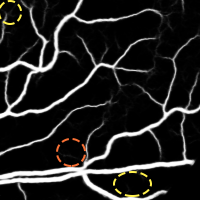}  \par\subcaption{}\label{fig:elast}
		\end{multicols}
		\vspace{-1cm}
		\begin{multicols}{4}
		\includegraphics[width=.25\textwidth]{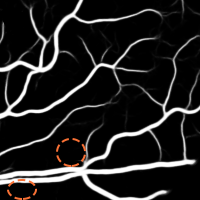}   \par\subcaption{}\label{fig:cons}
		\includegraphics[width=.25\textwidth]{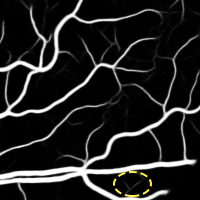}  \par\subcaption{}\label{fig:bs_simple}
		\includegraphics[width=.25\textwidth]{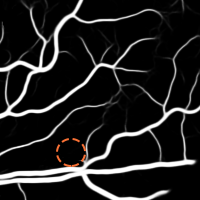}  \par\subcaption{}\label{fig:drop}
		\includegraphics[width=.25\textwidth]{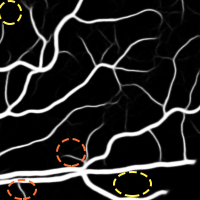}  \par\subcaption{}\label{fig:bs}
		\end{multicols}
		\vspace{-0.4cm}
		\caption {Effects of the components of the Base System on the probabilistic predictions : (a) Segmentation of the 1st human observer; (b) No Augmentation; (c)  Oversampling; (d) Elastic Samples; (e) Consecutive Rotations; (f) Simple Prediction; (g) Standard Dropout; (h) Base System. Yellow markings represent increase of false positives in relation to the Base System, while orange markings symbolize increase of false negatives.}
		\label{fig:components}
	\end{figure}

We started by evaluating our \textit{Base System} regarding data augmentation, prediction and regularization. The results of each variant are shown
in Table \ref{tab:ablation}, while the probabilistic predictions can be seen in Fig. \ref{fig:components}. All tests were performed under the same conditions, with the only source of variability being the component under study. 

\begin{table*}[h!]
\centering
\renewcommand{\arraystretch}{1}
\caption{Segmentation results of each test on the components of the Base System, on DRIVE. Bold values show the best score among all methods; underlined values represent metrics where the null hypothesis can be rejected ($p$-value $< 0.05$ when facing the Base System)}
\label{tab:ablation}
\scalebox{0.7}{
\begin{tabular}{lcccccccc}
\specialrule{.2em}{.1em}{.1em}
\multirow{2}{*}{\textbf{Method}} & \multicolumn{1}{l}{\multirow{2}{*}{\textbf{Dropout}}} & \multicolumn{3}{c}{\textbf{Patches per image}}                                                       & \multicolumn{1}{l}{\multirow{2}{*}{\textbf{Prediction}}} &  \multicolumn{1}{l}{\multirow{2}{*}{\textbf{Channels}}} & \multicolumn{1}{c}{\multirow{2}{*}{\textbf{\textit{Acc}}}} & \multicolumn{1}{c}{\multirow{2}{*}{\textbf{AUC}}} \\ \cline{3-5}
                        & \multicolumn{1}{l}{}                         & \multicolumn{1}{r}{Original} & \multicolumn{1}{c}{Rotated} & \multicolumn{1}{c}{Elastic} & \multicolumn{1}{l}{}                            & \multicolumn{1}{l}{}                     & \multicolumn{1}{l}{}                     \\ 
\specialrule{.12em}{.1em}{.1em}
No Augmentation & \multicolumn{1}{l}{Spatial} & \multicolumn{1}{r}{3000} & \multicolumn{1}{c}{-} & \multicolumn{1}{c}{-}  & \multicolumn{1}{l}{Multiple} & 1 & \multicolumn{1}{r}{\underline{0.9557}} & \multicolumn{1}{r}{\underline{0.9780}}  \\ 
Oversampling         & \multicolumn{1}{l}{Spatial} & \multicolumn{1}{r}{12000}& \multicolumn{1}{c}{-} & \multicolumn{1}{c}{-}  & \multicolumn{1}{l}{Multiple} & 1 & \multicolumn{1}{r}{\underline{0.9557}} & \multicolumn{1}{r}{\underline{0.9785}}  \\
Elastic Samples & \multicolumn{1}{l}{Spatial} & \multicolumn{1}{r}{3000} & \multicolumn{1}{c}{-} & \multicolumn{1}{c}{9000} & \multicolumn{1}{l}{Multiple} & 1 & \multicolumn{1}{r}{\underline{0.9558}} & \multicolumn{1}{r}{\underline{0.9785}}  \\
Consecutive Rotations   & \multicolumn{1}{l}{Spatial}   & \multicolumn{1}{r}{3000} & \multicolumn{1}{c}{9000} & \multicolumn{1}{c}{-}    & \multicolumn{1}{l}{Multiple} & 1 & \multicolumn{1}{r}{\underline{0.9560}} & \multicolumn{1}{r}{\underline{0.9795}}  \\
Simple Prediction       & \multicolumn{1}{l}{Spatial}   & \multicolumn{1}{r}{3000} & \multicolumn{1}{c}{9000}  & \multicolumn{1}{c}{-}    & \multicolumn{1}{l}{Simple}   & 1 & \multicolumn{1}{r}{\underline{0.9567}} & \multicolumn{1}{r}{\underline{0.9805}}  \\
Standard Dropout        & \multicolumn{1}{l}{Standard}  & \multicolumn{1}{r}{3000} & \multicolumn{1}{c}{9000}  & \multicolumn{1}{c}{-}    & \multicolumn{1}{l}{Multiple} & 1 & \multicolumn{1}{r}{\underline{0.9564}} & \multicolumn{1}{r}{\underline{0.9805}}  \\
\cline{1-9}
Base System              & \multicolumn{1}{l}{Spatial}   & \multicolumn{1}{r}{3000} & \multicolumn{1}{c}{9000}  & \multicolumn{1}{c}{-}   & \multicolumn{1}{l}{Multiple} & 1 & \multicolumn{1}{r}{\textbf{0.9570}} & \multicolumn{1}{r}{\textbf{0.9815}}  \\
\specialrule{.01em}{.1em}{.1em}
\end{tabular}}
\end{table*}

\subsubsection{Data Augmentation}
\label{sec:datacomponents}

To validate the procedure described in Section \ref{sec:PW}, we studied four alternatives. At first, we reduced the total number of patches per image to 3000, by not performing data augmentation -- \textit{No Augmentation}. Then, we increased the total number of patches to 12000 in three different ways. In the first case, we oversampled the image by extracting 9000 more original patches -- \textit{Oversampling}. In the second one, the remaining 9000 patches were artificially created by non-linearly deforming each patch, as described in \citep{oliveira} -- \textit{Elastic Samples}. Each set of 3000 elastic patches was obtained using a different $(\alpha, \sigma)$ combination: $(8, 1.5)$, $(16, 2.5)$, or $(32, 3)$. These values were found manually, ensuring that both the artificial samples and their respective annotations retain a consistent appearance. Finally, we used 9000 rotated patches, as in the \textit{Base System}, but we placed them consecutively (not randomly) -- \textit{Consecutive Rotations}.

Considering the \textit{Base System} as the reference, we can see that reducing the number of patches by four strongly deteriorated the results in terms of \textit{Acc} and AUC. Besides this, either when using original or elastic patches to keep the initial number of samples, the differences to the reference remained almost the same. Looking directly at Fig. \ref{fig:components}, we notice that these approaches favored the simultaneous appearance of FN and FP, with elastic patches leading to greater tortuosity in the detected vessel segments. Overall, this hints that the network benefited the most from the information encoded by the rotations. Another important note is related to the way those rotations were presented to the network. Recalling section \ref{sec:PW}, we have seen that some data augmentation strategies designed for FC-CNNs \citep{dielemancyclic, worrall2016harmonic} require the rotated versions to be arranged consecutively. The \textit{Consecutive Rotations} test hints that a deterministic proximity between the rotated patches is detrimental to the performance of the FCN,  with the network showing more difficulties in detecting vessel segments as can be seen in Fig. \ref{fig:components}.

When facing the randomly placed rotations used in the \textit{Base System}, all the alternatives were found to be prejudicial with statistical significance. 

\subsubsection{Multiple Prediction}

After verifying the benefits that the rotated patches brought to the model, we settled out to investigate whether these benefits could be extrapolated to the prediction. Our idea was to synchronously excite the network with the original patches and their rotations, performing the average of the outputs, as described in section \ref{sec:PW}.

Here, we compare the \textit{Base System}, which uses our multiple prediction scheme, with the \textit{Simple Prediction} variant. As can be seen in Table \ref{tab:ablation}, the \textit{Base System} performed better both in terms of \textit{Acc} and AUC. Moreover, comparing the predictions of both approaches (Fig. \ref{fig:components}), we notice that the multiple segmentation scheme makes the model less prone to FP, which is particularly important in medical applications. As a point of note, even if the changes in the mean values were slighter, statistically significant differences were found between the two methods.  

\subsubsection{Regularization}

Finally, we looked at the regularization of the model, by comparing the spatial dropout technique, described in section \ref{sec:reg}, with \textit{Standard Dropout}.

Still having the \textit{Base System} as a reference, we can see that the standard strategy led to a statistically significant drop in terms of \textit{Acc} and AUC (Table \ref{tab:ablation}). Overall, the predictions are quite similar, but the model seems to detect fewer vessel segments when the standard dropout is applied (Fig. \ref{fig:components}).

\subsection{Validation of the Stationary Wavelet Transform}
\label{sec:rswt}

\begin{figure*}[t!]
        
		\centering
		\begin{multicols}{4}
		\scalebox{1}{
		\hspace{-0.2cm} 
		\includegraphics[width=.25\textwidth]{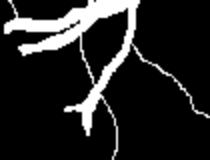}   
		\includegraphics[width=.25\textwidth]{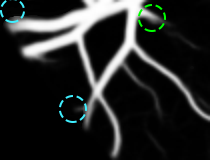}  
		\includegraphics[width=.25\textwidth]{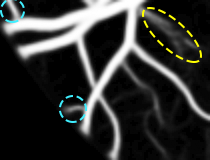}
		\includegraphics[width=.25\textwidth]{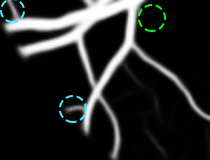}
		}
		\end{multicols}
		\vspace{-0.7cm}
		\begin{multicols}{4}
		\scalebox{1}{
		\hspace{-0.2cm} 
		\includegraphics[width=.25\textwidth]{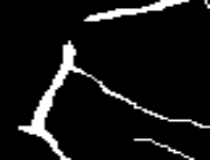}
		\includegraphics[width=.25\textwidth]{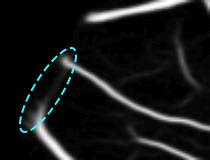}
		\includegraphics[width=.25\textwidth]{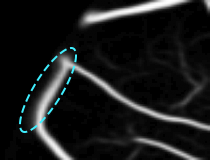}
		\includegraphics[width=.25\textwidth]{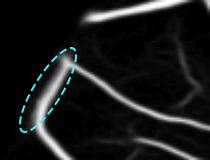}
		}
		\end{multicols}
		\vspace{-0.4cm}
		\small\text{(a) \hspace{.22\textwidth} (b)  \hspace{.22\textwidth} (c) \hspace{.22\textwidth} (d) }
		\caption {Effects of the inclusion of the SWT channels on the probabilistic predictions of two different patches: (a) Segmentation of the 1st human observer; (b) Base System; (c) BS + $d_1$; (d) BS + $d_2$ (best model). Green markings represent reduction of false positives in relation to the Base System; blue markings indicate reduction of false negatives; yellow markings symbolize increase of false positives.}
		\label{fig:exampleswavelets}
		
\end{figure*}

Having analyzed our \textit{Base System}, we evaluate the effects of incorporating the SWT into it. The results of each variant are shown in Table \ref{tab:ablationswt}, while the probabilistic predictions can be seen in Fig. \ref{fig:exampleswavelets}. The patches coming from the SWT were concatenated in the input, by varying the total number of input channels.  We resorted only to detail coefficients since the goal was to enhance image transitions. We started by concatenating the detail coefficients of the first level in the initial green channel input -- \textit{BS + $d_1$}. Then, we added those of the second level -- \textit{BS + $ d_1$ + $d_2$}. Finally, only the latter were kept -- \textit{BS + $d_2$}.

\begin{table*}[t]
\centering
\renewcommand{\arraystretch}{1.1}
\caption{Segmentation results of each  test on the inclusion of SWT, on  DRIVE. Bold values show the best score among all methods; underlined values represent metrics where the null hypothesis can be rejected ($p$-value $< 0.05$ when facing the Base System)}
\label{tab:ablationswt}
\scalebox{0.7}{
\begin{tabular}{lcccccccc}
\specialrule{.2em}{.1em}{.1em}
\multirow{2}{*}{\textbf{Method}} & \multicolumn{1}{l}{\multirow{2}{*}{\textbf{Dropout}}} & \multicolumn{3}{c}{\textbf{Patches per image}}                                                       & \multicolumn{1}{l}{\multirow{2}{*}{\textbf{Prediction}}} &  \multicolumn{1}{l}{\multirow{2}{*}{\textbf{Channels}}} & \multicolumn{1}{c}{\multirow{2}{*}{\textbf{\textit{Acc}}}} & \multicolumn{1}{c}{\multirow{2}{*}{\textbf{AUC}}} \\ \cline{3-5}
                        & \multicolumn{1}{l}{}                         & \multicolumn{1}{r}{Original} & \multicolumn{1}{c}{Rotated} & \multicolumn{1}{c}{Elastic} & \multicolumn{1}{l}{}                            & \multicolumn{1}{l}{}                     & \multicolumn{1}{l}{}                     \\ 
\specialrule{.12em}{.1em}{.1em}
Base System (BS)               & \multicolumn{1}{l}{Spatial}   & \multicolumn{1}{r}{3000} & \multicolumn{1}{c}{9000}  & \multicolumn{1}{c}{-}   & \multicolumn{1}{l}{Multiple} & 1 & \multicolumn{1}{r}{0.9570} & \multicolumn{1}{r}{0.9815}  \\
BS + $d_1$      & \multicolumn{1}{l}{Spatial}   & \multicolumn{1}{r}{3000} & \multicolumn{1}{c}{9000}  & \multicolumn{1}{c}{-}    & \multicolumn{1}{l}{Multiple}   & 4 & \multicolumn{1}{r}{0.9573} & \multicolumn{1}{r}{\underline{0.9818}}  \\
BS + $d_1$ + $d_2$      & \multicolumn{1}{l}{Spatial}   & \multicolumn{1}{r}{3000} & \multicolumn{1}{c}{9000}  & \multicolumn{1}{c}{-}    & \multicolumn{1}{l}{Multiple}   & 7 & \multicolumn{1}{r}{0.9574} & \multicolumn{1}{r}{\underline{0.9820}}  \\
\textbf{BS + $d_2$}   & \multicolumn{1}{l}{Spatial}   & \multicolumn{1}{r}{3000} & \multicolumn{1}{c}{9000}  & \multicolumn{1}{c}{-}    & \multicolumn{1}{l}{Multiple}   & 4 & \multicolumn{1}{r}{\underline{\textbf{0.9576}}} & \multicolumn{1}{r}{\underline{\textbf{0.9821}}}  \\
\specialrule{.01em}{.1em}{.1em}
\multicolumn{9}{l}{\begin{footnotesize}* For concision of notation,  $d_j$ represents the set consisting of $dV_j$, $dH_j$, and $dD_j$.\end{footnotesize}}
\end{tabular}}
\end{table*}

Analyzing the results of the tests performed, we notice that all the alternative strategies improved the performance of the \textit{Base System}, in terms of \textit{Acc} and AUC. This means that the use of features based on the wavelet decomposition, whose effectiveness for vessel segmentation is well-known \citep{soares,zhang2017}, is also beneficial when combined with a deep learning methodology. In particular, we see that the first level SWT coefficients used on \textit{BS + $d_1$} were less effective than those of the second level applied on \textit{BS + $d_2$}. The first level translates spectral information of higher frequencies; this seems to have induced more false positives, as can be seen in Fig. \ref{fig:exampleswavelets}. On the other hand, the second level SWT coefficients introduced statistically significant differences to the \textit{Base System}, in both \textit{Acc} and AUC, which seems to reinforce the idea that even deep learning methods can benefit from domain knowledge. In fact, they allowed to reduce the combination of false positives and false negatives as we can see in Fig. \ref{fig:exampleswavelets} as well. From now on, we will refer to the best model (\textit{BS + $ d_2$}) as \textit{Proposed}.

\subsection{Vessel Segmentation}

The binary segmentations were obtained by thresholding the probability maps at 0.5. We note that only pixels inside the field of view (FOV) were used in calculations.

The results for each database are shown in Table \ref{tab:soa}. The mean \textit{Acc} values obtained were higher than those of the second observer in all databases. The same occurred regarding \textit{Sp}, which reveals that the network presented few false positives, not signalling areas of injury, or spills, as vessels. In the DRIVE and CHASE\textunderscore DB1 databases, the mean \textit{Sn} surpassed that of the second observer, which shows that the network also rarely misclassified vessel pixels. In the STARE database, this tendency was not strong enough to level with the second observer, since he systematically marks vessels that the first one does not see. 

\begin{figure}[htbp!]
\centering
   \begin{subfigure}[h]{0.8\textwidth}
   \includegraphics[width=1\textwidth]{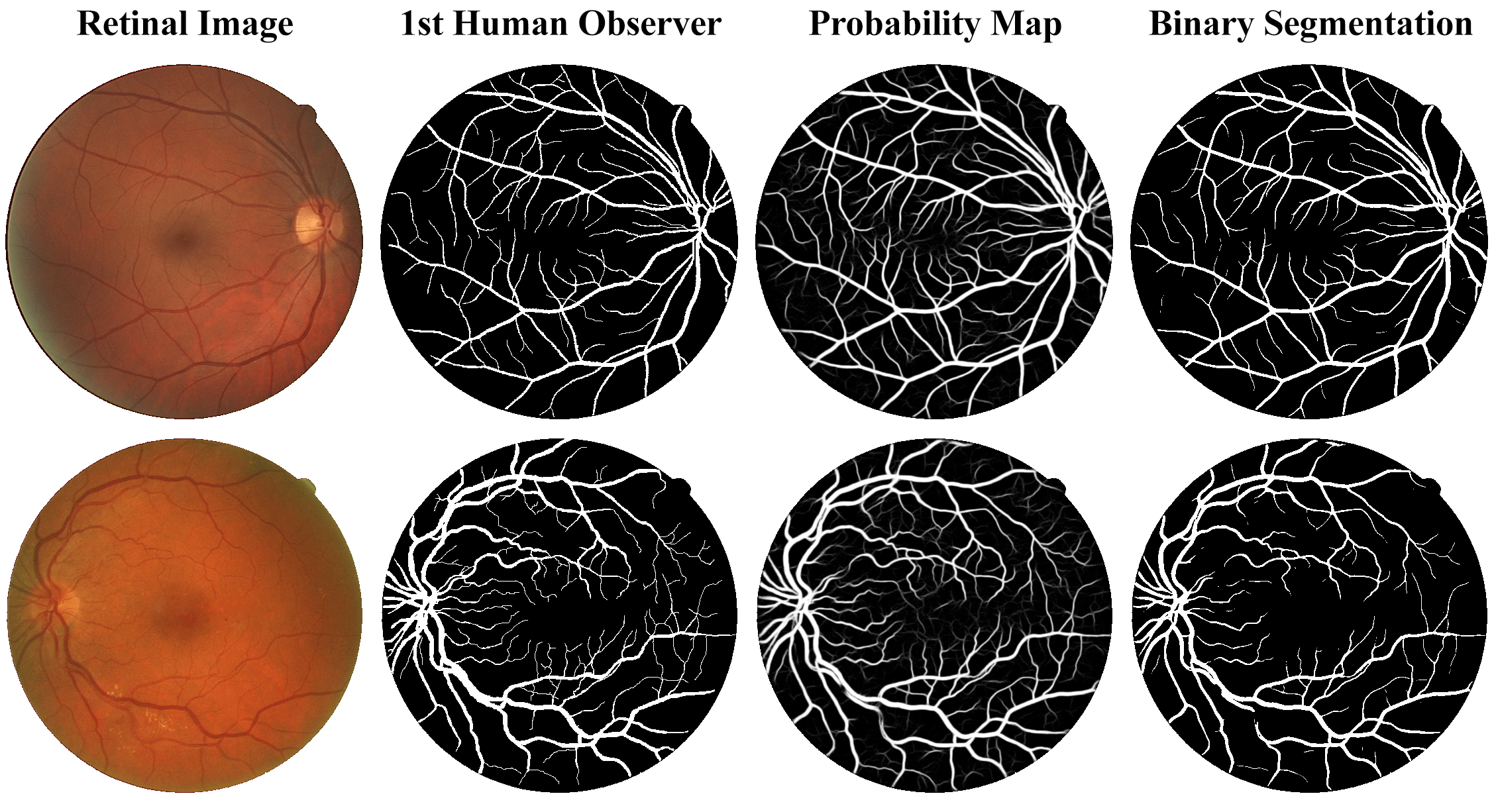}
   \subcaption{}
   \label{fig:segsdrive} 
   \end{subfigure}
   \begin{subfigure}[h]{0.8\textwidth}
   \includegraphics[width=1\textwidth]{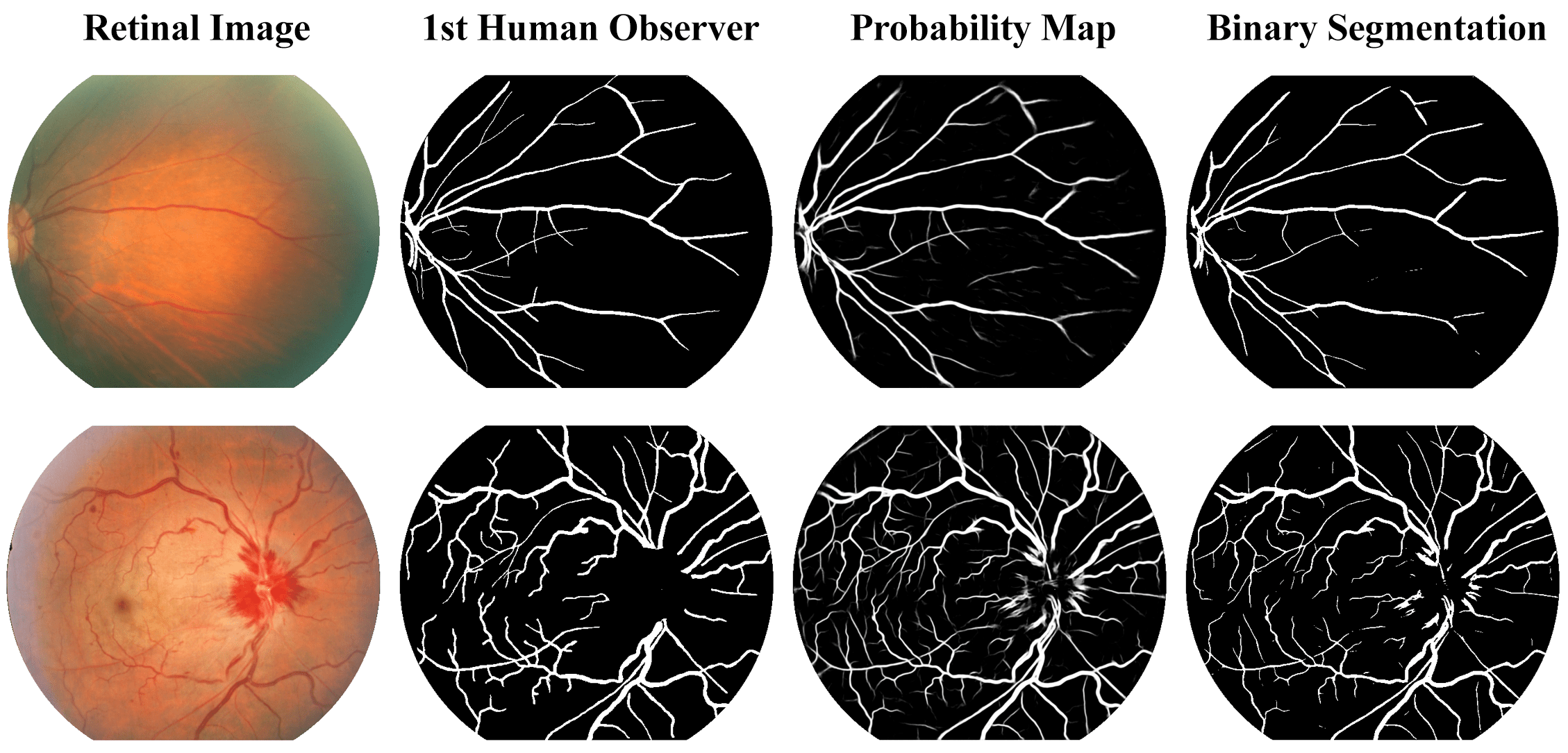}
   \subcaption{}
   \label{fig:segstare}
   \end{subfigure}
   \begin{subfigure}[h]{0.8\textwidth}
   \includegraphics[width=1\textwidth]{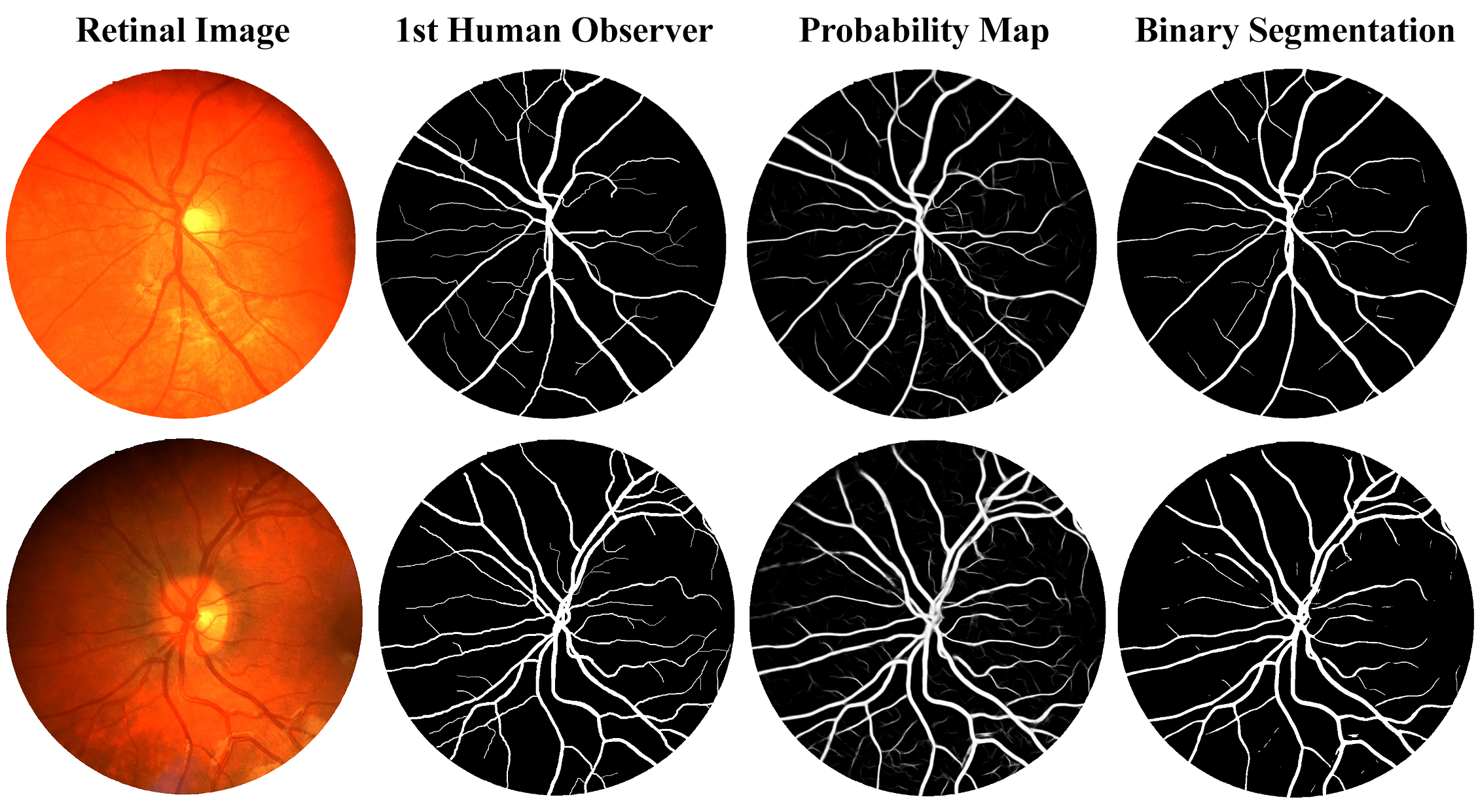}
   \subcaption{}
   \label{fig:segschase}
   \end{subfigure}
   \vspace{-0.1cm}
\caption{Segmentation examples for each database: (a) DRIVE; (b) STARE; (c) CHASE\textunderscore DB1. The first row shows the best case, while the second presents the worst one. }
\end{figure}

The \textit{Sn}, \textit{Sp}, \textit{Acc}, and AUC values of the best case for the DRIVE database were 0.9119, 0.9742, 0.9667, and 0.9903, while those of the worst case matched 0.7628, 0.9816, 0.9497, and 0.9786, respectively (Fig. \ref{fig:segsdrive}). For the STARE database, the values of the best case were 0.8527, 0.9936, 0.9837, and 0.9964, while those of the worst case matched 0.7231, 0.9827, 0.9503, and 0.9791 (Fig. \ref{fig:segstare}). Finally, for the CHASE\textunderscore DB1 database, the values of the best case were 0.8541, 0.9844, 0.9744, and 0.9909, while those of the worst case matched 0.8065, 0.9749, 0.9574, and 0.9808 (Fig. \ref{fig:segschase}). 

Regarding computational performance, the training of the network took about 4 hours, and each retinal image was fully segmented in approximately 2 seconds.

\begin{table*}[!h]
\caption{Segmentation results on the DRIVE, STARE, and CHASE\textunderscore DB1 databases. Bold values show the best score among all methods}
\begin{center}
\resizebox{\linewidth}{!}{
\renewcommand{\arraystretch}{1}
\scalebox{1}{
\begin{tabular}{cp{5.1cm}|c|cccc|cccc|cccc}
\specialrule{.2em}{.1em}{.1em}

&\multicolumn{1}{c}{}&        & \multicolumn{4}{c|}{\textbf{DRIVE}} & \multicolumn{4}{c|}{\textbf{STARE}} & \multicolumn{4}{c}{\textbf{CHASE\_DB1}} \\
\specialrule{.12em}{.1em}{.1em}
& \textbf{Method}   & \textbf{Year}   & \textit{\textbf{Sn}} & \textit{\textbf{Sp}}    & \textit{\textbf{Acc}} & \textbf{AUC}
                                      & \textit{\textbf{Sn}} & \textit{\textbf{Sp}}    & \textit{\textbf{Acc}} & \textbf{AUC}    
                                      & \textit{\textbf{Sn}} & \textit{\textbf{Sp}}    & \textit{\textbf{Acc}} & \textbf{AUC}   \\ 
\specialrule{.12em}{.1em}{.1em}

\multirow{7}[0]{*}{} 
& \multicolumn{1}{l|}{2nd human observer} & \multicolumn{1}{c|}{-}   & 0.7760      & 0.9725     & 0.9473 & -      & 0.8956 & 0.9381 & 0.9346 & -      & 0.7705     & 0.9778      & 0.9561      & - 
 \\ 
\midrule
          
\multirow{10}[0]{*}{\textbf{\shortstack {Unsupervised \\ methods}}} 
& \citet{hoover}           & 2000 & -      & -      & -      & -      & 0.6794 & 0.9560 & 0.9263 & -      & -      & -      & -      & - 
\\ 
& \citet{mendonça}         & 2006 & 0.7344 & 0.9764 & 0.9452 & -      & 0.6996 & 0.9730 & 0.9440 & -      & -      & -      & -      & -
\\
& \citet{lam}             & 2010 & -      & -      & 0.9472 & 0.9614 & -      & -      & 0.9567 & 0.9739 & -      & -      & -      & -
\\
& \citet{yintracking}      & 2012 & 0.6522 & 0.9710 & 0.9267 & -      & 0.7248 & 0.9666 & 0.9412 & -      & -      & -      & -      & -
\\
& \citet{nguyen}          & 2013 & -      & -      & 0.9407 & -      & -      & -      & 0.9324 & -      & -      & -      & -      & -
\\
& \citet{trucco}          & 2013 & -      & -      & 0.9461 & 0.9543 & -      & -      & 0.9521 & 0.9682 & -      & -      & -      & -
\\
& \citet{azzopardi}        & 2015 & 0.7655 & 0.9704 & 0.9442 & 0.9614 & 0.7716 & 0.9701 & 0.9497 & 0.9563 & 0.7585 & 0.9587 & 0.9387 & 0.9487 
\\ 
& \citet{roychowdhury2015} & 2015 & 0.7395 & 0.9782 & 0.9494 & 0.9672 & 0.7317 & 0.9842 & 0.9560 & 0.9673 & 0.7615 & 0.9575 & 0.9467 & 0.9623 
\\ 
& \citet{zhao}             & 2015 & 0.7420 & 0.9820 & 0.9540 & 0.8620 & 0.7800 & 0.9780 & 0.9560 & 0.8740 & -      & -      & -      & -     
\\
& \citet{zhang}            & 2016 & 0.7743 & 0.9725 & 0.9476 & 0.9636 & 0.7791 & 0.9758 & 0.9554 & 0.9748 & 0.7626 & 0.9661 & 0.9452 & 0.9606 
\\
& \citet{neto2017}         & 2017 & 0.7806 & 0.9629 & -      & -      & \textbf{0.8344} & 0.9443 & -      & -      & -      & -      & -    & -
\\
\midrule
  
\multirow{8}[0]{*}{\textbf{\shortstack {Supervised \\ methods}}} 
& \citet{niemeijer}        & 2004 & -      & -      & 0.9416 & 0.9294 & -      & -      & -      & -      & -      & -      & -      & - 
\\
& \citet{staal}            & 2004 & -      & -      & 0.9441 & 0.9520 & -      & -      & 0.9516 & 0.9614 & -      & -      & -      & - 
\\
& \citet{soares}           & 2006 & 0.7332 & 0.9782 & 0.9466 & 0.9614 & 0.7207 & 0.9747 & 0.9480 & 0.9671 & -      & -      & -      & - 
\\
& \citet{marin}            & 2011 & 0.7067 & 0.9801 & 0.9452 & 0.9588 & -      & -      & -      & -      & -      & -      & -      & -  
\\
& \citet{fraz}             & 2012 & 0.7406 & 0.9807 & 0.9480 & 0.9747 & 0.7548 & 0.9763 & 0.9534 & 0.9768 & 0.7224 & 0.9711 & 0.9469 & 0.9712 
\\
& \citet{roychowdhury2014} & 2015 & 0.7249 & \textbf{0.9830} & 0.9519 & 0.9620 & 0.7719 & 0.9726 & 0.9515 & 0.9688 & 0.7201 & 0.9824 & 0.9530 & 0.9532
\\
& \citet{strisciuglio2016} & 2016 & 0.7777 & 0.9702 & 0.9454 & 0.9597 & 0.8046 & 0.9710 & 0.9534 & 0.9638 & -      & -      & -      & - 
\\
& \citet{orlando}          & 2017 & 0.7897 & 0.9684 & 0.9454 & 0.9506 & 0.7680 & 0.9738 & 0.9519 & 0.9570 & 0.7565 & 0.9655 & 0.9467& 0.9478
\\
& \citet{zhang2017}        & 2017 & 0.7861 & 0.9712 & 0.9466 & 0.9703 & 0.7882 & 0.9729 & 0.9547 & 0.9740 & 0.7644 & 0.9716 & 0.9502     & 0.9706
\\
\midrule
\multirow{5}[0]{*}{\textbf{\shortstack {Deep Learning \\ methods}}} 
& \citet{melinscak}        & 2015 & 0.7276 & 0.9785 & 0.9466 & 0.9749 & -      & -      & -      & -      & -      & -      & -      & - 
\\
& \citet{feng}             & 2016 & 0.7569 & 0.9816 & 0.9527 & 0.9738 & 0.7726 & 0.9844 & 0.9628 & 0.9879 & 0.7507 & 0.9793 & 0.9581 & 0.9716 
\\
& \citet{liskowski}        & 2016 & 0.7520 & 0.9806 & 0.9515 & 0.9710 & 0.8145 & \textbf{0.9866} & \textbf{0.9696} & 0.9880 & -      & -      & -      & - \\

& \citet{fu}               & 2016 & 0.7603 & -      & 0.9523 & -      & 0.7412 & -      & 0.9585 & -      & 0.7130 & -      & 0.9489 & - 
\\

\cline{2-15}
& \textbf{Proposed}            & 2018 & \textbf{0.8039} & 0.9804 & \textbf{0.9576} & \textbf{0.9821} & 0.8315 & 0.9858 & 0.9694 & \textbf{0.9905} & \textbf{0.7779} & \textbf{0.9864} & \textbf{0.9653} & \textbf{0.9855} 
\\
\specialrule{.01em}{.1em}{.1em}
\\
\end{tabular}}}
\end{center}
\label{tab:soa}
\end{table*}

\subsection{Comparison with the State-of-the-art}

We compare our method with several other state-of-the-art methods in Table \ref{tab:soa}. 

We observe that our method obtained the highest \textit{Sn} in two of the three databases used (DRIVE and CHASE\textunderscore DB1), being very close to the best result in the third one (STARE), which indicates that it was capable of detecting many vessel pixels; also, considering the high value of \textit{Sp}, this was not obtained by increasing the number of false positives. This is significant, because training a classifier for segmenting retinal vessels becomes more difficult due to the unbalanced classes, where vessel pixels typically correspond only to 10\% of the image, and the detection of vessels without increasing false cases is essential for medical applications.

By jointly evaluating \textit{Sn}, \textit{Sp}, and \textit{Acc}, our method presented the best performance on both the DRIVE and CHASE\textunderscore DB1 databases.  In the DRIVE database, we outperformed all the other works regarding \textit{Sn} and \textit{Acc}. In terms of \textit{Sp}, we stood in sixth; however, knowing that \textit{Acc} combines information from \textit{Sn} and \textit{Sp}, we may conclude that the gain in true detections was much more significant than the inclusion of false detections. In other words, there is a notorious trade-off between \textit{Sn} and \textit{Sp}, with our method having a better balance. In the CHASE\textunderscore DB1 database, we ranked first in all metrics. Regarding the STARE database, we ranked second in all metrics, only behind \citet{neto2017} in \textit{Sn} and \citet{liskowski} in \textit{Sp} and \textit{Acc}. By comparing our work with \citep{neto2017}, we see that our \textit{Sp} is much higher so, given the trade-off abovementioned, the slight disadvantage in terms of \textit{Sn} sounds natural. In parallel, if we consider the second row of Fig. \ref{fig:segstare}, which shows our worst case, we notice that our method failed in rejecting small haemorrhagic blobs, which may have affected \textit{Sp}. This may be explained by our choice of using 5-fold cross-validation for training, which reduced the number of images with pathological signs, contrary to \citep{liskowski}, which used leave-one-out cross-validation.

Among the methods based on deep learning, only \citet{fu}  also used a multiscale FCN. If we compare this method with our \textit{Base System}, which did not use the extra channels from SWT, we verify that \citep{fu}  was surpassed in all available metrics. This means that the simpler structure of our encoder/decoder, when properly trained, was able to outperform a more elaborated architecture that joined a multiscale FCN with a CRF, and was trained end-to-end with a structured loss function. Comparing with all deep learning methods, we verify that our approach presented a much higher \textit{Sn} in all databases, and a much higher \textit{Acc} in DRIVE and CHASE\textunderscore DB1 databases. In DRIVE, which is perhaps the most used database in this area, we can see that before this work none of the previous best methods in \textit{Sn}, \textit{Sp}, and \textit{Acc}, was based on deep learning.

Contrary to \textit{Sn}, \textit{Sp}, and \textit{Acc}, the AUC score is not dependent on the threshold used to produce the final segmentations. Taking this into account, we note that our method achieved the highest AUC value in all databases. Also, to the best of our knowledge, we are the first to report AUC values above 0.98 for all databases.

\subsection{Cross-Training}

In a realistic situation, it is not feasible to retrain the model whenever new images have to be segmented. Additionally, a reliable method must successfully segment each image, even if the acquisition device belongs to a different manufacturer. That being said, robustness to the training set is crucial for the model to be of practical use. In this study, we did cross-training between the DRIVE and STARE databases, since they are the most widely used for this purpose \citep{soares,marin,fraz,roychowdhury2014,zhang2017,feng}. The results are shown in Table \ref{tab:cross}.

The \textit{Acc} average values went down from 0.9576 and 0.9694 to 0.9505 and 0.9597 for the DRIVE and STARE databases, respectively. This means decreases of about 0.7\% and 1.0\%, against the 0.2\% and 0.1\% of \citet{roychowdhury2014}. Regarding AUC, the results fell from 0.9821 and 0.9905 to 0.9748 and 0.9846, by the same order. This implies approximate reductions of 0.7\% and 0.6\%, against the 0.5\% and 1.1\% of \citet{fraz}.

As shown in Table \ref{tab:cross}, in absolute terms, we ranked first in all metrics, except \textit{Sn}, when training on STARE and testing on DRIVE. In the reverse case, something similar happened, but \textit{Sp} replaced \textit{Sn}. We have found that when training on STARE and testing on DRIVE, the network detects fewer thin vessels, so there was a drop in \textit{Sn}. In the reverse case, since the DRIVE database typically has more annotated thin vessels than STARE, \textit{Sn} rose considerably. 

\begin{table}[h]
\centering
\renewcommand{\arraystretch}{1}
\caption{Cross-training segmentation results between the DRIVE and STARE databases. Bold values show the best score among all methods}
\label{tab:cross}
\scalebox{0.7}{
\begin{tabular}{cp{4.8cm}cccc}
\specialrule{.2em}{.1em}{.1em}
\multicolumn{1}{c}{\begin{tabular}[c]{@{}c@{}} \textbf{Test set} \\ (\textbf{Training set})\end{tabular}}  & \multicolumn{1}{l}{\textbf{Method}} & \multicolumn{1}{c}{\textit{\textbf{Sn}}} & \multicolumn{1}{c}{\textit{\textbf{Sp}}} & \multicolumn{1}{c}{\textit{\textbf{Acc}}} & \multicolumn{1}{c}{\textbf{AUC}}\\

\specialrule{.12em}{.1em}{.1em}
\multirow{6}{*}{\begin{tabular}[c]{@{}c@{}} DRIVE \\ (STARE)\end{tabular}}    
& \citet{soares}                  & -                        & -                  & 0.9397              & -               \\
& \citet{marin}                    & -                        & -                  & 0.9448              & -               \\
& \citet{fraz}                      & 0.7242                   & 0.9792             & 0.9456              & 0.9697          \\
& \citet{roychowdhury2014}  & -                        & -                  & 0.9494              & -               \\
& \citet{feng}                        & \textbf{0.7273}          & 0.9810             & 0.9486              & 0.9677          \\
& \citet{zhang2017}                & -                        & -                  & 0.9447              & 0.9593          \\
\cline{2-6}
& \textbf{Proposed}                     & 0.6706                   & \textbf{0.9916}    & \textbf{0.9505}     & \textbf{0.9748} \\
\midrule
\multirow{6}{*}{\begin{tabular}[c]{@{}c@{}} STARE \\ (DRIVE)\end{tabular}}    
& \citet{soares}                  & -                & -                  & 0.9327             & -                  \\
& \citet{marin}                    & -                & -                  & 0.9526             & -                  \\
& \citet{fraz}                      & 0.7010           & 0.9770             & 0.9495             & 0.9660             \\
& \citet{roychowdhury2014}  & -                & -                  & 0.9510             & -                  \\
& \citet{feng}                        & 0.7027           & \textbf{0.9828}    & 0.9545             & 0.9671             \\
& \citet{zhang2017}                & -                & -                  & 0.9488             & 0.9676             \\
\cline{2-6}
& \textbf{Proposed}         & \textbf{0.8453}  & 0.9726             & \textbf{0.9597}   & \textbf{0.9846}    \\
\specialrule{.01em}{.1em}{.1em}
\end{tabular}}
\end{table}

\subsection{Robustness to the Inter-Rater Variability}
\label{inter-rater}

The manual annotations provided by physicians are crucial for training and testing supervised methods. Experience may help experts refine their segmentations, but two different observers will always have their own way of approaching the task. There is not always agreement between the experts since some of them systematically see more vessels than others. Moreover, even when the vessel is clear, differences in the estimation of its caliber may arise. For these reasons, inter-rater variability is always present in the evaluation process. Here, we discuss how this variability affects the obtained results.

All the previous results reported in this work were obtained using the annotations of the first human observer as the gold standard for training and testing. In Table \ref{tab:robust}, otherwise, the gold standard, for testing, were the annotations of the second human observer. Having these as ground truth, we then compared both the first observer and the model, for each database. Regarding the tests of statistical significance, we aimed to reject the following null hypothesis: the results of the \textit{Proposed} method are not statistically different from those of the first human observer.

\begin{table}[h!]
\centering
\renewcommand{\arraystretch}{1}
\caption{Segmentation results on the DRIVE, STARE, and CHASE\textunderscore DB1 databases, using the annotations of the 2nd human observer as gold standard for testing. We also evaluate the 1st human observer in relation to the 2nd. Bold values show the best score between the two methods; underlined values represent metrics where the null hypothesis can be rejected ($p$-value $< 0.05$ when they are confronted)}
\label{tab:robust}
\scalebox{0.7}{
\begin{tabular}{cp{3.5cm}ccc}
\specialrule{.2em}{.1em}{.1em}
\multicolumn{1}{c}{\textbf{Test set}}  & \multicolumn{1}{c}{\textbf{Method}} & \multicolumn{1}{c}{\textit{\textbf{Sn}}} & \multicolumn{1}{c}{\textit{\textbf{Sp}}} & \multicolumn{1}{c}{\textit{\textbf{Acc}}} \\
\specialrule{.12em}{.1em}{.1em}
\multirow{2}{*}{DRIVE}    
& 1st human observer        & 0.8066                       & 0.9674                        & 0.9473                        \\
& \textbf{Proposed}         & \underline{\textbf{0.8405}}  & \underline{\textbf{0.9814}}   & \underline{\textbf{0.9639}}   \\
\midrule

\multirow{2}{*}{STARE}                  
& 1st human observer        & \textbf{0.6439}              & 0.9883                        & 0.9346                        \\
& \textbf{Proposed}         & 0.6329                       & \underline{\textbf{0.9924}}   & \textbf{0.9365}               \\
\midrule

\multirow{2}{*}{CHASE\textunderscore DB1}                  
& 1st human observer        & \textbf{0.7974}              & 0.9736                        & 0.9561                         \\
& \textbf{Proposed}         & 0.7731                     & \underline{\textbf{0.9813}}   & \underline{\textbf{0.9600}}    \\
\specialrule{.01em}{.1em}{.1em}
\end{tabular}}
\end{table}

\newpage

As shown in Table \ref{tab:robust}, if we consider the second observer as the gold standard, the model outperformed the first observer regarding \textit{Sp} and \textit{Acc}, in the STARE and CHASE\textunderscore DB1 databases. According to the definition of \textit{Sp}, this means that it introduced fewer false positives than the first observer himself (in relation to the second one). In other words, some of the pixels marked as background by the model, and rejected by the first observer, ended up being validated by the second one. In the DRIVE database, the model performed better than the first observer in terms of \textit{Sn}, \textit{Sp}, and \textit{Acc}. This shows that apart from having less false positives, it also presented less false negatives than the first observer. That is, even though the model was trained according to the first observer, it could identify vessel segments that only the second observer saw. Together, the previous results suggest that even when the model was trained according to the first observer, it seemed to be more consistent than him when another observer was considered. To put it another way, the differences established between two independent human observers appeared to be higher than those presented by the model when it was evaluated against a new reference. This reveals a consistency which may be unreachable to a human who always oscillates from day to day and can be affected by fatigue or stress.

\begin{figure}[!h]
		\centering
		\begin{multicols}{3}
		\includegraphics[width=.25\textwidth, scale=0.6]{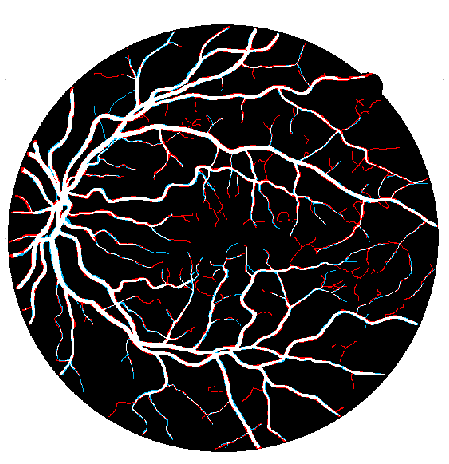} \par
		\null\vspace{0.01cm}\includegraphics[width=.25\textwidth, scale=0.6]{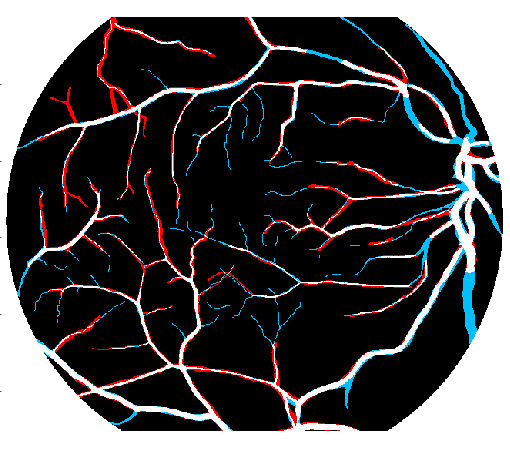} \par
		\includegraphics[width=.25\textwidth, scale=0.6]{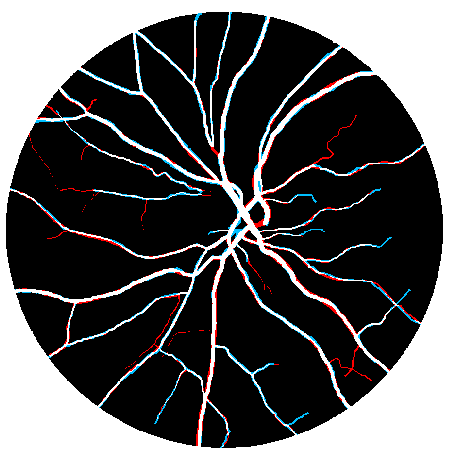} \par
		\end{multicols}

		\begin{multicols}{3}
		\includegraphics[width=.25\textwidth, scale=0.6]{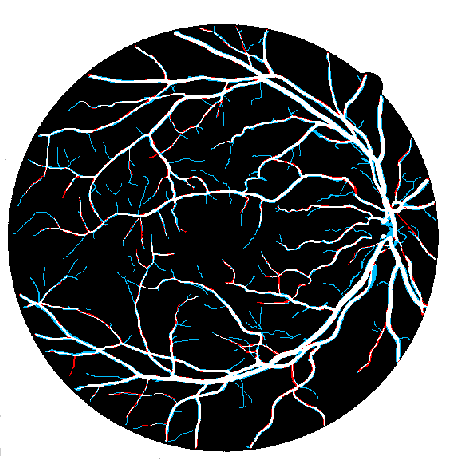} \par\subcaption{}
		\null\vspace{0.03cm}\includegraphics[width=.25\textwidth, scale=0.6]{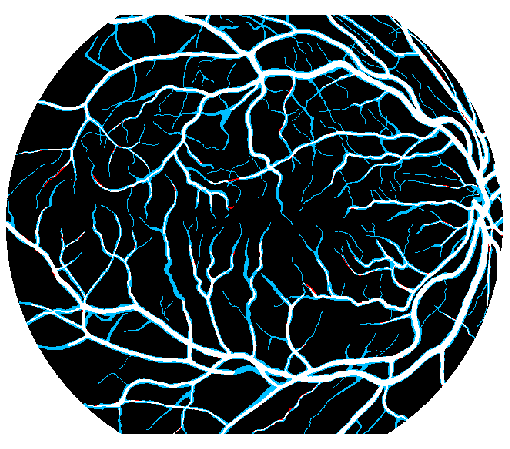} \par\subcaption{}
		\includegraphics[width=.25\textwidth, scale=0.6]{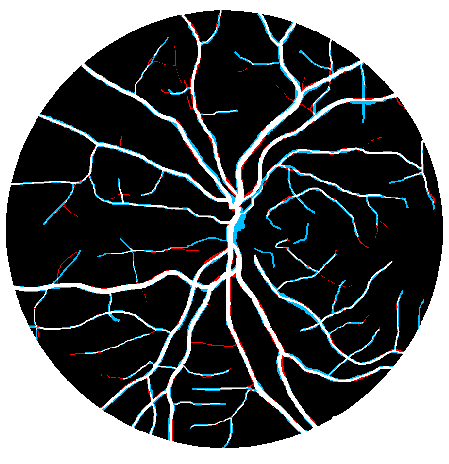} \par\subcaption{}
		\end{multicols}
		
		\caption {Comparison between the annotations of the 1st and 2nd human observers for each database: (a) DRIVE; (b) STARE; and (c) CHASE\textunderscore DB1. The first row shows the comparison for the case where the model improves the most being evaluated according to the 2nd observer (with the \textit{Acc} values for the model/ 1st observer being, from (a) to (c): 0.9692/0.9467, 0.9647/0.9527, and 0.9643/0.9613); the second row refers to the case where it most worsens (with the \textit{Acc} values becoming: 0.9582/0.9446, 0.9028/0.8968, and 0.9492/0.9480). Each color represents a different situation: black -- both mark as background; white -- both mark as vessel; red -- only the 1st observer marks as vessel; blue -- only the 2nd observer marks as vessel.}
		\label{fig:gt}
\end{figure}
 
Comparing Tables \ref{tab:soa} and \ref{tab:robust}, we can further analyze how the model behaved when we changed the  reference from the first to the second observer. For the STARE and CHASE\textunderscore DB1 databases, the average \textit{Acc} of the model fell from 0.9694 and 0.9653 to 0.9365 and 0.9600, respectively. This means approximate decreases of 3.4\% and 0.5\%, which sound natural since it was trained according to the first observer. For the STARE database, in particular, the decrease was higher since the second observer systematically marks more vessels than the first one. The DRIVE database case was surprising. The average \textit{Acc} rose from 0.9576 to 0.9639, which corresponds to an approximate increase of 0.7\%. This means that despite being trained according to the first observer, the model was closer to the second one.

Except for the STARE database, in both the remaining databases there is no clear difference between the marking pattern of the experts. That is, we can see cases where the first observer marks vessels that the second one does not see and vice versa. This may be deduced from Fig. \ref{fig:gt}, where we compare the annotations of both observers, for each database. We believe that this noticeable discordance may become a limiting factor for future improvements since the current results of this field are already extremely accurate.

\section{Conclusions}
\label{sec:conclusions}

In this paper, we proposed a novel FCN-based method for retinal vessel segmentation. We used rotation operations for data augmentation and introduced a new way to use the information they provide, during training, to strengthen the prediction. Moreover, we investigated the addition of new channels into a FCN through the SWT decomposition. This allowed to improve the performance in terms of Acc and AUC, reducing the combination of FP and FN and suggesting that deep learning methods still leave room for domain knowledge. Results on three publicly available databases showed that we are competitive with state-of-the-art methods. In terms of \textit{Acc}, we rank first on the DRIVE and CHASE\textunderscore DB1 databases, while being second in STARE. In terms of AUC, we lead on all of them. 

Our method proved to be robust to the training set and to the inter-rater variability, which shows its potential for real-world application on screening and diagnostic systems. Due to its simple convolutional nature, and by using an efficient GPU implementation, it also proves to be fast. Fully segmenting a retinal image takes approximately 2 seconds.

Based on our vessel segmentation results, it may be possible to extract different biomarkers from retinal images and apply them for clinical purposes. Also, the proposed method can be further applied on other types of medical images for segmenting elongated structures. 

Although the proposed approach shows very god performance, there are still some aspects that can be further explored in the future. In this work, we showed that extra features were beneficial for the segmentation performance. Hence, domain knowledge may improve Deep Learning-based models. So, as long term research, we will explore further the combination of Deep Learning with domain knowledge. In retinal imaging, we believe that other wavelet families, such as curvelets, may also prove useful. Finally, the amount of training data may limit the capacity of the CNNs. Hence, we expect that as access to medical data becomes easier, we will be able to design different architectures.

\section*{ACKNOWLEDGMENT}
This work is supported by FCT with the reference project UID/EEA/04436/2013, by FEDER funds through the COMPETE 2020 – Programa Operacional Competitividade e Internacionaliza\c{c}\~ao (POCI) with the reference project POCI-01-0145-FEDER-006941. 

\bibliographystyle{apalike}
\bibliography{references}

\begin{thebibliography}{}

\bibitem[Abr{\`a}moff et~al., 2010]{abramoff_review}
Abr{\`a}moff, M.~D., Garvin, M.~K., and Sonka, M. (2010).
\newblock Retinal imaging and image analysis.
\newblock {\em IEEE Rev. Biomed. Eng.}, 3:169--208.

\bibitem[Azzopardi et~al., 2015]{azzopardi}
Azzopardi, G., Strisciuglio, N., Vento, M., and Petkov, N. (2015).
\newblock Trainable cosfire filters for vessel delineation with application to
  retinal images.
\newblock {\em Med. Image. Anal.}, 19(1):46--57.

\bibitem[Badrinarayanan et~al., 2015]{segnet}
Badrinarayanan, V., Kendall, A., and Cipolla, R. (2015).
\newblock Segnet: A deep convolutional encoder-decoder architecture for image
  segmentation.
\newblock {\em arXiv preprint arXiv:1511.00561}.

\bibitem[Bengio et~al., 2013]{bengio}
Bengio, Y., Courville, A., and Vincent, P. (2013).
\newblock Representation learning: A review and new perspectives.
\newblock {\em IEEE Trans. Pattern Anal. Mach. Intell.}, 35(8):1798--1828.

\bibitem[Dai et~al., 2015]{boxsup}
Dai, J., He, K., and Sun, J. (2015).
\newblock Boxsup: Exploiting bounding boxes to supervise convolutional networks
  for semantic segmentation.
\newblock In {\em Proceedings of the IEEE International Conference on Computer
  Vision}, pages 1635--1643.

\bibitem[Dieleman et~al., 2016]{dielemancyclic}
Dieleman, S., De~Fauw, J., and Kavukcuoglu, K. (2016).
\newblock Exploiting cyclic symmetry in convolutional neural networks.
\newblock {\em arXiv preprint arXiv:1602.02660}.

\bibitem[Dieleman et~al., 2015]{dieleman}
Dieleman, S., Willett, K.~W., and Dambre, J. (2015).
\newblock Rotation-invariant convolutional neural networks for galaxy
  morphology prediction.
\newblock {\em Mon. Not. R. Astron. Soc.}, 450(2):1441--1459.

\bibitem[Duan et~al., 2017]{duansar}
Duan, Y., Liu, F., Jiao, L., Zhao, P., and Zhang, L. (2017).
\newblock Sar image segmentation based on convolutional-wavelet neural network
  and markov random field.
\newblock {\em Pattern Recognition}, 64:255--267.

\bibitem[Eigen and Fergus, 2015]{predicting}
Eigen, D. and Fergus, R. (2015).
\newblock Predicting depth, surface normals and semantic labels with a common
  multi- convolutional architecture.
\newblock In {\em Proceedings of the IEEE International Conference on Computer
  Vision}, pages 2650--2658.

\bibitem[Fraz et~al., 2012a]{fraz_review}
Fraz, M.~M., Remagnino, P., Hoppe, A., Uyyanonvara, B., Rudnicka, A.~R., Owen,
  C.~G., and Barman, S.~A. (2012a).
\newblock Blood vessel segmentation methodologies in retinal images -- a
  survey.
\newblock {\em Comput. Methods Programs Biomed.}, 108(1):407--433.

\bibitem[Fraz et~al., 2012b]{fraz}
Fraz, M.~M., Remagnino, P., Hoppe, A., Uyyanonvara, B., Rudnicka, A.~R., Owen,
  C.~G., and Barman, S.~A. (2012b).
\newblock An ensemble classification-based approach applied to retinal blood
  vessel segmentation.
\newblock {\em IEEE Trans. Biomed. Eng.}, 59(9):2538--2548.

\bibitem[Fu et~al., 2016]{fu}
Fu, H., Xu, Y., Lin, S., Wong, D. W.~K., and Liu, J. (2016).
\newblock Deepvessel: Retinal vessel segmentation via deep learning and
  conditional random field.
\newblock In {\em MICCAI}, pages 132--139. Springer.

\bibitem[Glorot and Bengio, 2010]{xavier}
Glorot, X. and Bengio, Y. (2010).
\newblock Understanding the difficulty of training deep feedforward neural
  networks.
\newblock In {\em Aistats}, volume~9, pages 249--256.

\bibitem[Guo et~al., 2017]{guo}
Guo, T., Mousavi, H.~S., Vu, T.~H., and Monga, V. (2017).
\newblock Deep wavelet prediction for image super-resolution.
\newblock In {\em The IEEE Conference on Computer Vision and Pattern
  Recognition (CVPR) Workshops}.

\bibitem[Haar, 1910]{haar1910}
Haar, A. (1910).
\newblock Zur theorie der orthogonalen funktionensysteme.
\newblock {\em Mathematische Annalen}, 69(3):331--371.

\bibitem[Hariharan et~al., 2015]{hypercolumns}
Hariharan, B., Arbel{\'a}ez, P., Girshick, R., and Malik, J. (2015).
\newblock Hypercolumns for object segmentation and fine-grained localization.
\newblock In {\em Proceedings of the IEEE Conference on Computer Vision and
  Pattern Recognition}, pages 447--456.

\bibitem[Holschneider et~al., 1990]{holschneider}
Holschneider, M., Kronland-Martinet, R., Morlet, J., and Tchamitchian, P.
  (1990).
\newblock A real-time algorithm for signal analysis with the help of the
  wavelet transform.
\newblock In {\em Wavelets}, pages 286--297. Springer.

\bibitem[Hoover et~al., 2000]{hoover}
Hoover, A., Kouznetsova, V., and Goldbaum, M. (2000).
\newblock Locating blood vessels in retinal images by piecewise threshold
  probing of a matched filter response.
\newblock {\em IEEE Trans. Med. Imag.}, 19(3):203--210.

\bibitem[Krizhevsky et~al., 2012]{krizhevsky}
Krizhevsky, A., Sutskever, I., and Hinton, G.~E. (2012).
\newblock Imagenet classification with deep convolutional neural networks.
\newblock In {\em Adv. Neural Inf. Process Syst.}, pages 1097--1105.

\bibitem[Lam et~al., 2010]{lam}
Lam, B.~S., Gao, Y., and Liew, A. W.-C. (2010).
\newblock General retinal vessel segmentation using regularization-based
  multiconcavity modeling.
\newblock {\em IEEE Trans. Med. Imag.}, 29(7):1369--1381.

\bibitem[LeCun et~al., 2015]{lecun}
LeCun, Y., Bengio, Y., and Hinton, G. (2015).
\newblock Deep learning.
\newblock {\em Nature}, 521(7553):436--444.

\bibitem[LeCun et~al., 1990]{lecun1990}
LeCun, Y., Boser, B.~E., Denker, J.~S., Henderson, D., Howard, R.~E., Hubbard,
  W.~E., and Jackel, L.~D. (1990).
\newblock Handwritten digit recognition with a back-propagation network.
\newblock In {\em Adv. Neural. Inf. Process. Syst.}, pages 396--404.

\bibitem[Li et~al., 2016]{feng}
Li, Q., Feng, B., Xie, L., Liang, P., Zhang, H., and Wang, T. (2016).
\newblock A cross-modality learning approach for vessel segmentation in retinal
  images.
\newblock {\em IEEE Trans. Med. Imag.}, 35(1):109--118.

\bibitem[Liskowski and Krawiec, 2016]{liskowski}
Liskowski, P. and Krawiec, K. (2016).
\newblock Segmenting retinal blood vessels with deep neural networks.
\newblock {\em IEEE Trans. Med. Imag.}, 35(11):2369--2380.

\bibitem[Liu and Sun, 1993]{liu}
Liu, I. and Sun, Y. (1993).
\newblock Recursive tracking of vascular networks in angiograms based on the
  detection-deletion scheme.
\newblock {\em IEEE Trans. Med. Imag.}, 12(2):334--341.

\bibitem[Long et~al., 2015]{long}
Long, J., Shelhamer, E., and Darrell, T. (2015).
\newblock Fully convolutional networks for semantic segmentation.
\newblock In {\em CVPR}, pages 3431--3440.

\bibitem[Mar{\'\i}n et~al., 2011]{marin}
Mar{\'\i}n, D., Aquino, A., Geg{\'u}ndez-Arias, M.~E., and Bravo, J.~M. (2011).
\newblock A new supervised method for blood vessel segmentation in retinal
  images by using gray-level and moment invariants-based features.
\newblock {\em IEEE Trans. Med. Imag.}, 30(1):146--158.

\bibitem[Melin{\v{s}}{\v{c}}ak et~al., 2015]{melinscak}
Melin{\v{s}}{\v{c}}ak, M., Prenta{\v{s}}i{\'c}, P., and Lon{\v{c}}ari{\'c}, S.
  (2015).
\newblock Retinal vessel segmentation using deep neural networks.
\newblock In {\em VISAPP}.

\bibitem[Mendonca and Campilho, 2006]{mendonça}
Mendonca, A.~M. and Campilho, A. (2006).
\newblock Segmentation of retinal blood vessels by combining the detection of
  centerlines and morphological reconstruction.
\newblock {\em IEEE Trans. Med. Imag.}, 25(9):1200--1213.

\bibitem[Nair and Hinton, 2010]{nair}
Nair, V. and Hinton, G.~E. (2010).
\newblock Rectified linear units improve restricted boltzmann machines.
\newblock In {\em ICML}, pages 807--814.

\bibitem[Nesterov, 1983]{nesterov}
Nesterov, Y. (1983).
\newblock A method of solving a convex programming problem with convergence
  rate o (1/k2).
\newblock In {\em Soviet Mathematics Doklady}, volume~27, pages 372--376.

\bibitem[Neto et~al., 2017]{neto2017}
Neto, L.~C., Ramalho, G.~L., Neto, J. F.~R., Veras, R.~M., and Medeiros, F.~N.
  (2017).
\newblock An unsupervised coarse-to-fine algorithm for blood vessel
  segmentation in fundus images.
\newblock {\em Expert Systems with Applications}, 78:182--192.

\bibitem[Nguyen et~al., 2013]{nguyen}
Nguyen, U.~T., Bhuiyan, A., Park, L.~A., and Ramamohanarao, K. (2013).
\newblock An effective retinal blood vessel segmentation method using multi-
  line detection.
\newblock {\em Pattern Recogn.}, 46(3):703--715.

\bibitem[Niemeijer et~al., 2004]{niemeijer}
Niemeijer, M., Staal, J., van Ginneken, B., Loog, M., and Abramoff, M.~D.
  (2004).
\newblock Comparative study of retinal vessel segmentation methods on a new
  publicly available database.
\newblock In {\em Medical Imaging}, volume 5370, pages 648--656. SPIE.

\bibitem[Noh et~al., 2015]{learning}
Noh, H., Hong, S., and Han, B. (2015).
\newblock Learning deconvolution network for semantic segmentation.
\newblock In {\em Proceedings of the IEEE International Conference on Computer
  Vision}, pages 1520--1528.

\bibitem[Oliveira et~al., 2017]{oliveira}
Oliveira, A., Pereira, S., and Silva, C.~A. (2017).
\newblock Augmenting data when training a cnn for retinal vessel segmentation:
  How to warp?
\newblock In {\em ENBENG}, pages 1--4. IEEE.

\bibitem[Orlando et~al., 2017]{orlando}
Orlando, J.~I., Prokofyeva, E., and Blaschko, M.~B. (2017).
\newblock A discriminatively trained fully connected conditional random field
  model for blood vessel segmentation in fundus images.
\newblock {\em IEEE Trans. Biomed. Eng.}, 64(1):16--27.

\bibitem[Owen et~al., 2009]{owen}
Owen, C.~G., Rudnicka, A.~R., Mullen, R., Barman, S.~A., Monekosso, D.,
  Whincup, P.~H., Ng, J., and Paterson, C. (2009).
\newblock Measuring retinal vessel tortuosity in 10-year-old children:
  validation of the computer-assisted image analysis of the retina (caiar)
  program.
\newblock {\em Invest. Ophthalmol. Vis. Sci.}, 50(5):2004--2010.

\bibitem[Patton et~al., 2006]{patton}
Patton, N., Aslam, T.~M., MacGillivray, T., Deary, I.~J., Dhillon, B.,
  Eikelboom, R.~H., Yogesan, K., and Constable, I.~J. (2006).
\newblock Retinal image analysis: concepts, applications and potential.
\newblock {\em Prog. Retin. Eye Res.}, 25(1):99--127.

\bibitem[Ronneberger et~al., 2015]{ronneberger}
Ronneberger, O., Fischer, P., and Brox, T. (2015).
\newblock U-net: Convolutional networks for biomedical image segmentation.
\newblock In {\em MICCAI}, pages 234--241. Springer.

\bibitem[Roychowdhury et~al., 2015a]{roychowdhury2014}
Roychowdhury, S., Koozekanani, D.~D., and Parhi, K.~K. (2015a).
\newblock Blood vessel segmentation of fundus images by major vessel extraction
  and subimage classification.
\newblock {\em IEEE journal of biomedical and health informatics},
  19(3):1118--1128.

\bibitem[Roychowdhury et~al., 2015b]{roychowdhury2015}
Roychowdhury, S., Koozekanani, D.~D., and Parhi, K.~K. (2015b).
\newblock Iterative vessel segmentation of fundus images.
\newblock {\em IEEE Transactions on Biomedical Engineering}, 62(7):1738--1749.

\bibitem[Simonyan and Zisserman, 2014]{simonyan}
Simonyan, K. and Zisserman, A. (2014).
\newblock Very deep convolutional networks for large- image recognition.
\newblock {\em arXiv preprint arXiv:1409.1556}.

\bibitem[Soares et~al., 2006]{soares}
Soares, J.~V., Leandro, J.~J., Cesar, R.~M., Jelinek, H.~F., and Cree, M.~J.
  (2006).
\newblock Retinal vessel segmentation using the 2d gabor wavelet and supervised
  classification.
\newblock {\em IEEE Trans. Med. Imag.}, 25(9):1214--1222.

\bibitem[Srivastava et~al., 2014]{srivastava}
Srivastava, N., Hinton, G.~E., Krizhevsky, A., Sutskever, I., and
  Salakhutdinov, R. (2014).
\newblock Dropout: a simple way to prevent neural networks from overfitting.
\newblock {\em J. Mach. Learn. Res.}, 15(1):1929--1958.

\bibitem[Staal et~al., 2004]{staal}
Staal, J., Abr{\`a}moff, M.~D., Niemeijer, M., Viergever, M.~A., and van
  Ginneken, B. (2004).
\newblock Ridge-based vessel segmentation in color images of the retina.
\newblock {\em IEEE Trans. Med. Imag.}, 23(4):501--509.

\bibitem[Strisciuglio et~al., 2016]{strisciuglio2016}
Strisciuglio, N., Azzopardi, G., Vento, M., and Petkov, N. (2016).
\newblock Supervised vessel delineation in retinal fundus images with the
  automatic selection of b-cosfire filters.
\newblock {\em Machine Vision and Applications}, 27(8):1137--1149.

\bibitem[Tompson et~al., 2015]{tompson}
Tompson, J., Goroshin, R., Jain, A., LeCun, Y., and Bregler, C. (2015).
\newblock Efficient object localization using convolutional networks.
\newblock In {\em CVPR}, pages 648--656.

\bibitem[Wang et~al., 2013]{trucco}
Wang, Y., Ji, G., Lin, P., and Trucco, E. (2013).
\newblock Retinal vessel segmentation using multiwavelet kernels and multi
  hierarchical decomposition.
\newblock {\em Pattern Recogn.}, 46(8):2117--2133.

\bibitem[Wilcoxon, 1945]{wilcoxon}
Wilcoxon, F. (1945).
\newblock Individual comparisons by ranking methods.
\newblock {\em Biometrics bulletin}, 1(6):80--83.

\bibitem[Worrall et~al., 2016]{worrall2016harmonic}
Worrall, D.~E., Garbin, S.~J., Turmukhambetov, D., and Brostow, G.~J. (2016).
\newblock Harmonic networks: Deep translation and rotation equivariance.
\newblock {\em arXiv preprint arXiv:1612.04642}.

\bibitem[Yin et~al., 2012]{yintracking}
Yin, Y., Adel, M., and Bourennane, S. (2012).
\newblock Retinal vessel segmentation using a probabilistic tracking method.
\newblock {\em Pattern Recognition}, 45(4):1235--1244.

\bibitem[Zhang et~al., 2017]{zhang2017}
Zhang, J., Chen, Y., Bekkers, E., Wang, M., Dashtbozorg, B., and ter
  Haar~Romeny, B.~M. (2017).
\newblock Retinal vessel delineation using a brain-inspired wavelet transform
  and random forest.
\newblock {\em Pattern Recogn.}, 69:107--123.

\bibitem[Zhang et~al., 2016]{zhang}
Zhang, J., Dashtbozorg, B., Bekkers, E., Pluim, J.~P., Duits, R., and ter
  Haar~Romeny, B.~M. (2016).
\newblock Robust retinal vessel segmentation via locally adaptive derivative
  frames in orientation scores.
\newblock {\em IEEE Trans. Med. Imag.}, 35(12):2631--2644.

\bibitem[Zhao et~al., 2015]{zhao}
Zhao, Y., Rada, L., Chen, K., Harding, S.~P., and Zheng, Y. (2015).
\newblock Automated vessel segmentation using infinite perimeter active contour
  model with hybrid region information with application to retinal images.
\newblock {\em IEEE Trans. Med. Imag.}, 34(9):1797--1807.

\end{thebibliography}

\end{document}